\documentclass[a4paper,fleqn,usenatbib]{mnras}

\usepackage{newtxtext,newtxmath}

\usepackage[T1]{fontenc}
\usepackage{ae,aecompl}

\usepackage{graphicx}	
\usepackage{amsmath}	
\usepackage{amssymb}	
\usepackage{aas_macros}

\newcommand{\kms}{km\,s$^{-1}$}

\title[SNhunt151]{SNhunt151: an explosive event inside a dense cocoon} 

\author[Elias-Rosa et al.]{N. Elias-Rosa$^{1,2}$\thanks{E-mail: nancy.elias@oapd.inaf.it}, 
S. Benetti$^{1}$, 
E. Cappellaro$^{1}$, 
A. Pastorello$^{1}$, 
G. Terreran$^{1}$,
\newauthor
A. Morales-Garoffolo$^{2}$,
S. C. Howerton$^{3}$,
S. Valenti$^{4}$,  
E. Kankare$^{5}$, 
A.~J. Drake$^{6}$,
\newauthor
S. G. Djorgovski$^{6}$,
L. Tomasella$^{1}$,
L. Tartaglia$^{7,1}$, 
T. Kangas$^{8}$,
P. Ochner$^{1}$, 
\newauthor
A. V. Filippenko$^{9,10}$,
F. Ciabattari$^{11}$,
S. Geier$^{12,13}$,
D.~A. Howell$^{14,15}$,
J. Isern$^{2}$,
\newauthor
S. Leonini$^{16}$,
G. Pignata$^{17,18}$,
M. Turatto$^{1}$
\\
\\
$^{1}$ INAF -- Osservatorio Astronomico di Padova, vicolo dell'Osservatorio 5, Padova I-35122, Italy \\
$^{2}$ Department of Applied Physics, University of C\'adiz, Campus of Puerto Real, 11510 C\'adiz, Spain\\
$^{3}$ 1401 South A, Arkansas City, KS 67005, USA\\
$^{4}$ Department of Physics, University of California, Davis, CA 95616, USA\\
$^{5}$ Astrophysics Research Centre, School of Mathematics and Physics, Queen's University Belfast, Belfast BT7 1NN, United Kingdom\\
$^{6}$ Astronomy Department, California Institute of Technology, Pasadena, CA 91125, USA\\
$^{7}$ Texas Tech University, Physics Department, Box 41051, Lubbock, TX 79409-1051, USA\\
$^{8}$ Tuorla Observatory, Department of Physics and Astronomy, University of Turku, V\"ais\"al\"antie 20, FI-21500 Piikki\"o, Finland\\
$^{9}$ Department of Astronomy, University of California, Berkeley, CA 94720-3411, USA\\
$^{10}$ Miller Senior Fellow, Miller Institute for Basic Research in Science, University of California, Berkeley, CA 94720, USA\\
$^{11}$ Osservatorio Astronomico di Monte Agliale, Via Cune Motrone, 55023 Borgo a Mozzano, Lucca, Italy\\ 
$^{12}$ Gran Telescopio Canarias (GRANTECAN), Cuesta de San Jos\'e s/n, E-38712, Bre\~na Baja, La Palma, Spain\\
$^{13}$ Instituto de Astrof\'{i}sica de Canarias, V\'{i}a L\'actea s/n, E-38200, La Laguna, Tenerife, Spain\\
$^{14}$ Las Cumbres Observatory, 6740 Cortona Drive, Suite 102, Goleta, CA 93117, USA\\
$^{15}$ Department of Physics, University of California, Santa Barbara, Broida Hall, Mail Code 9530, Santa Barbara, CA 93106-9530, USA\\
$^{16}$ Osservatorio Astronomico Provinciale di Montarrenti, S.S. 73 Ponente, I-53018 Sovicille, Siena, Italy\\
$^{17}$ Departamento de Ciencias F\'{i}sicas, Universidad Andres Bello, Avda. Rep\'ublica 252, Santiago, 8320000, Chile\\
$^{18}$ Millennium Institute of Astrophysics (MAS), Nuncio Monse\~nor S\'otero Sanz 100, Providencia, Santiago, Chile\\
}

\date{Accepted XXX. Received YYY; in original form ZZZ}

\pubyear{2017}

\begin{document}
\label{firstpage}
\pagerange{\pageref{firstpage}--\pageref{lastpage}}
\maketitle

\begin{abstract}
SNhunt151 was initially classified as a supernova (SN) impostor (nonterminal outburst of a massive star). It exhibited a slow increase in luminosity, lasting about 450~d, followed by a major brightening that reaches  $M_V \approx -18$ mag. No source is detected to $M_V \gtrsim -13$ mag in archival images at the position of SNhunt151 before the slow rise. Low-to-mid-resolution optical spectra obtained during the pronounced brightening show very little evolution, being dominated at all times by multicomponent Balmer emission lines, a signature of interaction between the material ejected in the new outburst and the pre-existing circumstellar medium. We also analyzed mid-infrared images from the {\sl Spitzer Space Telescope}, detecting a source at the transient position in 2014 and 2015. Overall, SNhunt151 is spectroscopically a Type IIn SN, somewhat similar to SN~2009ip. However, there are also some differences, such as a slow pre-discovery rise, a relatively broad light-curve peak showing a longer rise time ($\sim 50$ d) and a slower decline, along with a negligible change in the temperature around the peak ($T \leq 10^4$ K). We suggest that SNhunt151 is the result of an outburst, or a SN explosion, within a dense circumstellar nebula, similar to those embedding some luminous blue variables like $\eta$ Carinae and originating from past mass-loss events.

\end{abstract}



\section{Introduction}\label{intro}

Massive stars can lose mass via steady winds, binary interaction, or as a consequence of dramatic eruptions, generating dense and structured circumstellar cocoons. When these stars explode as supernovae (SNe), the ejecta interact with a pre-existing, optically thick circumstellar medium (CSM), producing collisionless shocks (see, e.g., \citealt{chevalier12} or \citealt{ofek13a}) that convert the ejecta kinetic energy into radiation. This interaction strongly affects the SN spectroscopic and photometric display. The study of these observables gives us important clues to the progenitors' mass-loss rates, hence constraining the late evolution of massive stars (see, e.g., \citealt{smith14c}).

Early-time spectra of H-rich interacting SNe (Type IIn) are characterized by a blue continuum with narrow hydrogen emission lines, with full width at half-maximum (FWHM) velocities up to about 2000 \kms. These are believed to arise from the  ionization of the pre-existing CSM because of its interaction with the SN ejecta. The narrow lines may have Lorentzian wings caused by electron scattering. However, relatively broad components  (FWHM $\approx 10^4$ \kms) can  also be produced by fast-moving SN ejecta and are detectable when the CSM is optically thin (see, e.g., \citealt{terreran16}). Interacting SNe display a large variety of spectral line components (having different widths and profiles) and photometric parameters (SNe with both slowly and rapidly declining light curves). These observational differences indicate a substantial diversity in the energetics, as well as in the configuration of the progenitor star and its environment at the moment of explosion.

The typical SN signatures can be masked at all phases if the SN ejecta interact with dense CSM. In particular, the elements produced during the explosive nucleosynthesis (including O, C, Mg) are not frequently detected in SNe~IIn. 

A growing number of interacting SNe preceded by major optical outbursts weeks to years before the SN explosion \citep{ofek14} has been observed. These pre-SN outbursts are often called ``SN impostors'' (e.g., SN 1997bs,  \citealt{vandyk00}; or the general discussions by \citealt{smith11d} and \citealt{vandyk_rev12}) because they mimic in terms of energetic and spectral appearance a real SN explosion, though the star survives the burst. The discrimination between SN impostors and SNe~IIn can be challenging. When outbursts of massive stars herald terminal SN explosions, the instabilities are presumably related to physical processes occurring when the star approaches the end of its evolution (see, e.g., \citealt{quataert12,smith14b,fuller17}). Well-studied examples include SN~2009ip (\citealt{smith10,foley11,pastorello13,fraser13,mauerhan13a,margutti14,fraser15,graham17}), SN~2010mc \citep{ofek13b,smith14}, LSQ13zm \citep{tartaglia16}, SN~2015bh \citep{ofek16,eliasrosa16,thone17}, and SN~2016bdu \citep{pastorello17}.

\begin{figure*}
\centering
\includegraphics[width=2.1\columnwidth]{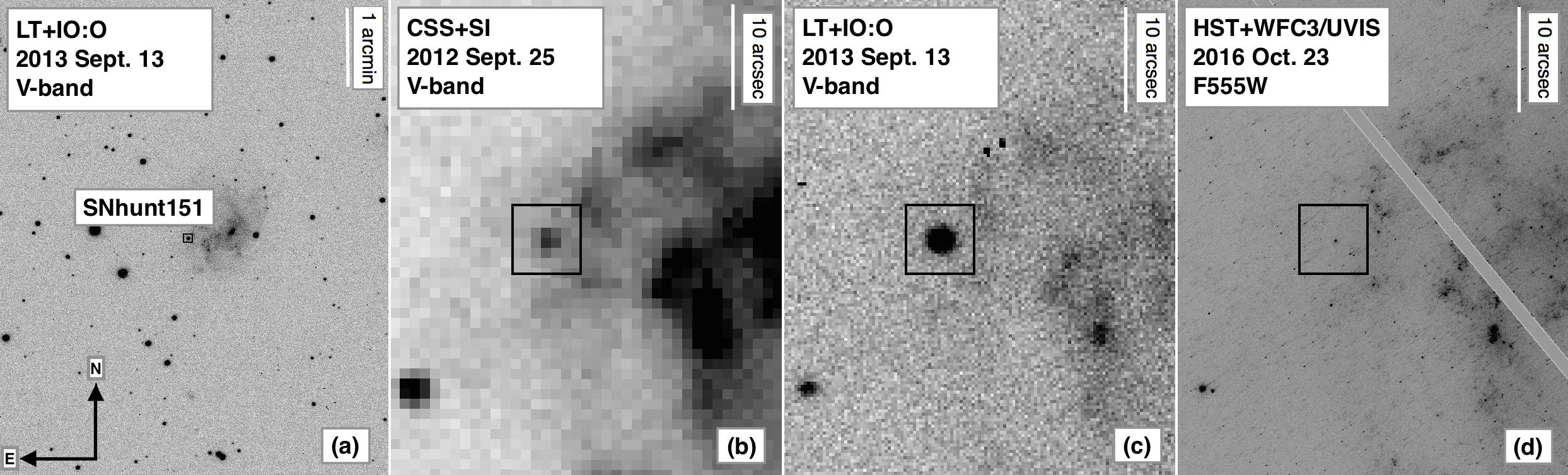}
 \caption{$V$- and $F555W$-band images of SNhunt151 in UGC~3165 obtained with the Liverpool Telescope+IO:O on 2013 September 13 {\it (a,c)}, the Cassegrain reflector+SI on 2012 September 25 {\it (b)}, and {\sl HST}+WFC3/UVIS on 2016 October 23 {\it (d)}. The location of the transient is indicated.}
\label{fig_transient}
\end{figure*}

In this context, SNhunt151 (also known as PSN J04472985+2358555, XM07BW, and CSS131025:044730+235856) is an interesting object. It was discovered by the Catalina Real-Time Transient Survey (CRTS; \citealt{drake09})\footnote{\url{http://crts.caltech.edu.}} on 2012 September 25.41 (UT dates are used hereafter) at $\alpha=04^{\rm h} 47^{\rm m} 30{\fs}00, \delta=+23^{\circ} 58\arcmin 56{\farcs}0$ (J2000.0; Fig. \ref{fig_transient}). The object, hosted in the spiral galaxy UGC~3165, had an unfiltered magnitude of $20.7$. The transient was reobserved, on 2013 August 30.17, showing a significant increase in its brightness, up to $17.6$ mag. 

The presence of an H$\alpha$ emission line in the spectrum of 2013 September 02.21, showing a complex profile composed of broad and narrow components \citep{pastorello_atel13,pastorello_atel13_2}, is reminiscent of SN~2009ip. After the initial brightening, a further increase (by about 2 mag) in the luminosity was reported about 50~d later \citep{eliasrosa_atel13}.

In this manuscript, we present our study of SNhunt151. Basic information on the SN host galaxy, UGC~3165, is found in Section \ref{SNhostgx}. Results of the photometric and spectroscopic monitoring campaign of SNhunt151 are presented in Sections \ref{SNph} and \ref{SNspec}. A discussion and brief summary follow in Section \ref{SNdiscus}.

%
\section{Host Galaxy, Distance, and Reddening}\label{SNhostgx}
UGC~3165 is classified as an irregular galaxy (IAm\footnote{NED, NASA/IPAC Extragalactic Database; http://nedwww.ipac.caltech.edu/.}). From the average recession velocity of the galaxy corrected for Local Group infall into the Virgo cluster \citep{mould00} $v_{\rm Vir} = 3783 \pm 3$ \kms\ ($z = 0.0126$), and assuming H$_0$ = 73 \kms\ Mpc$^{-1}$, we derive a distance of $51.8 \pm 3.6$ Mpc ($\mu = 33.57 \pm 0.15$ mag). This distance will be adopted throughout this paper.

The Milky Way extinction along the line of sight of the transient is quite large, $A_{V, \rm MW} = 1.88$ mag (NED; \citealt{schlafly11}). Additionally, the higher-resolution spectra of the transient (taken on 2013 September 2 and 2013 November 1; see Table \ref{table_spec}) exhibit a blended Na\,{\sc i}~D feature at the recession velocity of the host galaxy, with an average equivalent width (EW) of $0.26 \pm 0.02$ \AA. Using the relations given by \cite{turatto03} and \cite{poznanski12}, and assuming $R_V = 3.1$ \citep{cardelli89}, we derive $A_{V, \rm host} \approx 0.09 \pm 0.02$ mag. Thus, we adopt $A_{V, \rm tot} = 1.97 \pm 0.19$ mag\footnote{We consider a conservative uncertainty in the Milky Way extinction of about 10\%.}($E(B-V)_{\rm tot} = 0.64 \pm 0.06$ mag) as the total extinction toward SNhunt151. We caution that a large dispersion is observed  in the EW (Na\,{\sc i}~D) versus $E(B-V)$ plane when a large SN sample is considered. For this reason, the derived extinction estimate cannot be considered secure (see, e.g., \citealt{eliasrosa07,poznanski11,phillips13}).
%

\section{Photometry}\label{SNph}

\subsection{Ground-Based Observations}\label{Ground}

Optical {\it BVRI} (Johnson-Cousins system), {\it ugriz}  (Sloan system), and near-infrared (NIR) {\it JHK} images of SNhunt151 were taken using a large number of observing facilities, listed in Table \ref{table_setup}. After discovery, the transient was observed for about 14 months with a gap of $\sim 5$ months when it was behind the Sun. Our data include 48 epochs taken before the discovery, starting on December 2003 [upper limit from the Sloan Digital Sky Survey (SDSS) DR9 \footnote{http://www.sdss.org.}] up to March 2012. These data were taken mainly by amateur astronomers. One epoch was secured using the Ultraviolet and Optical Telescope on board the {\it Swift} satellite on 2013 September 05.65. However, only upper limits were estimated for the {\it UVW2} ($>20.9$ mag) and {\it UVW1} ($>20.0$ mag) bands ({\sc Vegamag}). 

Photometric observations were preprocessed following the standard recipe in {\sc iraf}\footnote{{\sc iraf} is distributed by the National Optical Astronomy Observatory, which is operated by the Associated Universities for Research in Astronomy, Inc., under cooperative agreement with the National Science Foundation.} for CCD images (trimming, overscan, bias, and flat-field corrections). For the NIR exposures, we also applied an illumination correction and background-sky subtraction. For later epochs, multiple exposures obtained in the same night and filter were combined to improve the signal-to-noise ratio (S/N). The SN magnitudes were measured using point-spread-function (PSF) fitting with SNOoPy\footnote{SNOoPy is a package for SN photometry developed by E. Cappellaro. A package description can be found at http://sngroup.oapd.inaf.it/snoopy.html.}. No template subtraction was performed since no major differences were found when comparing the photometry obtained through both PSF fitting and template subtraction. The  data derived with the two methods are in agreement to within 0.1 mag.

In order to calibrate the instrumental magnitudes to standard photometric systems, we used the zero points and colour terms measured through reference SDSS stars in the field of SNhunt151. The Johnson-Cousins magnitudes of these reference stars were calculated from the Sloan magnitudes following the conversions of \cite{chonis08}. For the NIR photometry, we used the Two Micron All Sky Survey (2MASS) catalog as reference for the calibration. Once the local sequence was calibrated, we proceeded to calibrate our transient, even for non-photometric nights.

Unfiltered magnitudes from amateurs were scaled to Johnson-Cousins $V$- or $R$-band magnitudes, depending on the wavelength peak efficiency of the detectors. 
Uncertainty estimates were obtained through artificial-star experiments. This is combined (in quadrature) with the PSF fit error returned by DAOPHOT and the propagated errors from the photometric calibration.

The final calibrated SN magnitudes are listed in Tables \ref{table_JCph} and \ref{table_SLph} for the optical bands, and in Table \ref{table_NIRph} for the NIR range.

\subsection{Space-Based Observations}\label{Ground}

The SNhunt151 site was also observed by the {\sl Hubble Space Telescope\/} ({\sl HST}) with the Ultraviolet-Visible (UVIS; $\sim 0.04''$ pixel$^{-1}$) Channel of the Wide-Field Camera 3 (WFC3) in the $F555W$ and $F814W$ bands. These images were taken on 2016 October 23 as part of the SN snapshot survey GO-14668 (PI: A. V. Filippenko), 1113.1 days after the $V$-band peak of SNhunt151's Event B (see below).

 A source was detected at the transient position in all images with root-mean-square uncertainties $< 0.01''$ (Fig. \ref{fig_transient}{\it (d)}), through comparison with ground-based, post-discovery Copernico Telescope+AFOSC and NOT+ALFOSC images taken on 2013 August 31 (seeing $1.16''$) and 2014 August 24 (seeing $0.96''$). 
 
The magnitudes of the source in \textsc{vegamag} were obtained using Dolphot\footnote{Dolphot is a stellar photometry package that was adapted from {\sc HSTphot} for general use \citep{dolphin00} with the individual WFC3 pre-mosaic frames. We used the WFC3 module of v2.0, updated 2016 April; \url{http://americano.dolphinsim.com/dolphot/.}}. We assume that the $F555W$ and $F814W$ bandpasses correspond approximately to Johnson-Cousins $V$ and $I$, respectively; thus, we include these measurements in Table~\ref{table_JCph}.

\subsection{Light Curves}\label{LC}

\begin{figure*}
\centering
\includegraphics[width=1.4\columnwidth]{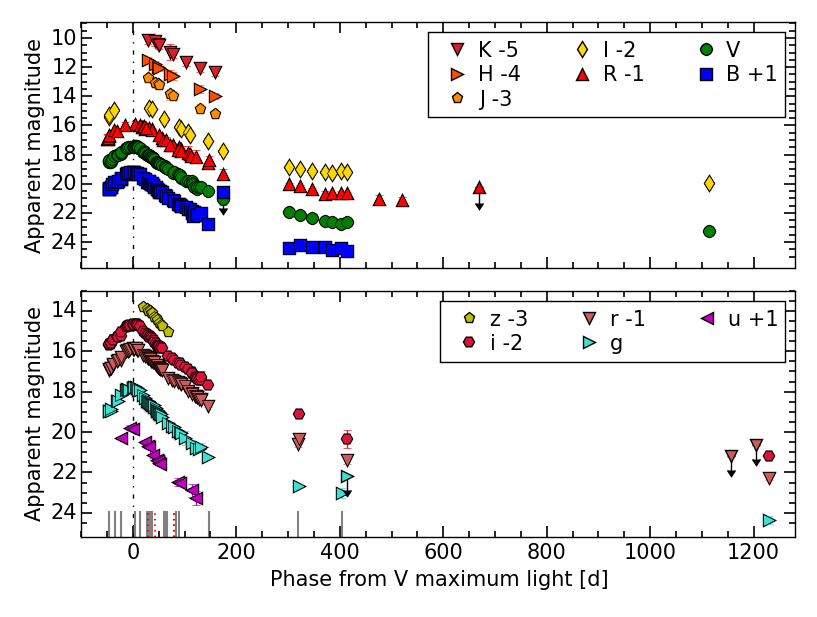}
\caption{Optical and NIR light curves of SNhunt151. Upper limits are indicated by a symbol with an arrow. The solid and dotted marks on the abscissa indicate the phases at which optical and NIR spectra were obtained, respectively. The dot-dashed vertical line indicates the $V$-band maximum light of SNhunt151. The light curves have been shifted for clarity by the amounts indicated in the legend. The uncertainties for most data points are smaller than the plotted symbols. A colour version of this figure can be found in the online journal.}
\label{fig_ph}
\end{figure*}

The $ugri$ and $BVRIJHK$ light curves are shown in Figure~\ref{fig_ph}, relative to the $V$ maximum date on 2013 October 06.3, or MJD 56571.8 (estimated by low-order polynomial fitting; in the following we will adopt this epoch as the reference). As we can see, SNhunt151 was discovered well before maximum light. The light curves also show a relatively fast decline after maximum, similar to those of the rapidly declining SNe~II (e.g., \citealt{anderson14}). Optical light curves of SNhunt151 show post-maximum decline rates between 1.8 and 2.8 mag (100 d)$^{-1}$, and those in the NIR have a slightly slower decline ($\sim 1.8$ mag (100 d)$^{-1}$). At late times ($> 300$ d), SNhunt151 decreases in brightness by $\sim 5$ mag with respect to maximum light, and shows a slower decline ($<1.3$ mag (100 d)$^{-1}$). Table \ref{table_max} summarises the magnitude and epochs at maximum light, as well as the decline rates of SNhunt151.

\begin{figure*}
\centering
\includegraphics[width=1.6\columnwidth]{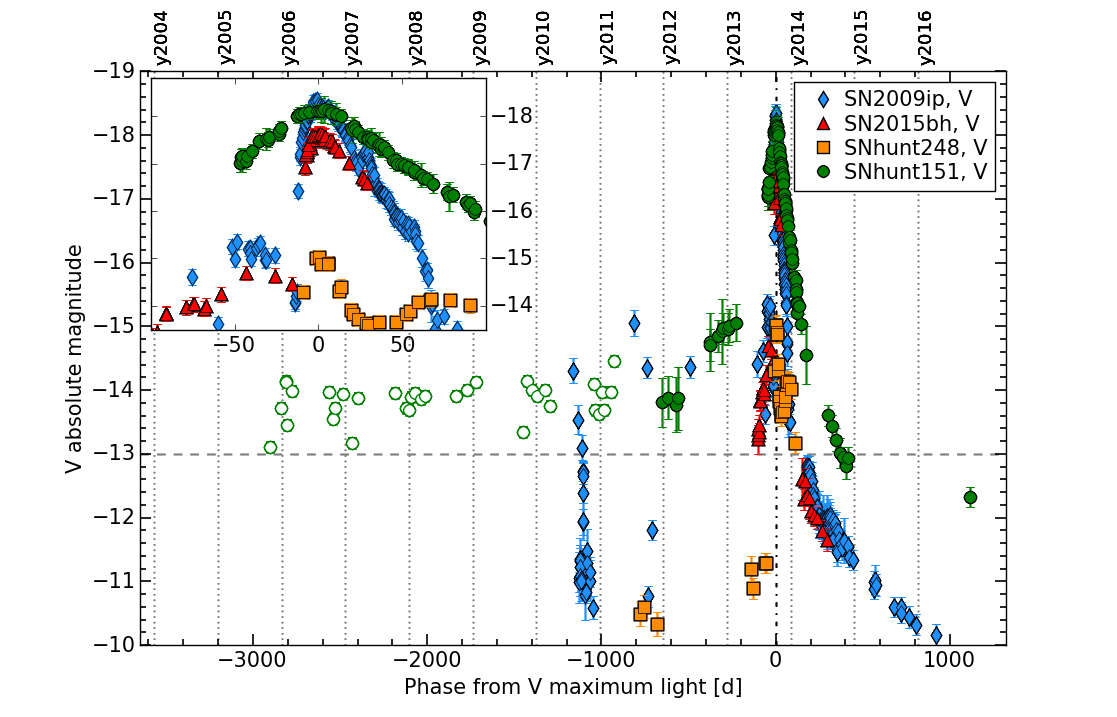}
\caption{Historical absolute $V$ light curve of SNhunt151, shown along with those of SN~2009ip, SN~2015bh, and SNhunt248. Upper limits for SNhunt151 are indicated by empty circles with arrows. The dashed horizontal line indicates the boundary of $-13$ mag, while the dot-dashed vertical line indicates the $V$-band maximum light of SNhunt151. Note that the first epoch of SNhunt151 is an approximation from Sloan magnitudes. A colour version of this figure can be found in the online journal.}
\label{fig_abs}
\end{figure*}

The SNhunt151 site was observed several times over almost a decade before the transient discovery (see Table \ref{table_JCph} and Figure \ref{fig_abs}, and Figure \ref{fig_transient}{\it (b)}). Until the end of 2011, we did not detect a source at the position of the transient, at an absolute magnitude brighter than $-13$ mag. However, from December 2011 we began to detect some flux, followed by a slow luminosity rise (labelled Event A) of about 1.6 mag in 450 days ($\sim 0.2$ mag (100 d)$^{-1}$), corresponding to a rise from $-13.8$ mag to $-15$ mag. Note that this rise may also consist of multiple distinct outbursts, since the observational campaign was interrupted twice. After August 2013, the transient brightness showed a much steeper increase, reaching a peak of $-18.10$ mag in $V$ (labelled Event B), consistent with the typical magnitudes at peak of most Type IIn SNe (between $-16$ and $-19$ mag; \citealt{kiewe12}), or the Event B of the SN 2009ip-like family of transients ($M_{r/R} \lesssim -18$ mag; Pastorello et al. 2017, submitted).

The $V$ absolute magnitude light curve of SNhunt151 is compared in Figure \ref{fig_abs} with those of SN~2009ip \citep{maza09,smith10,foley11,pastorello13,mauerhan13a,fraser13}, SN~2015bh \citep{eliasrosa16}, and the SN~impostor SNhunt248 \citep{kankare15}. For these objects we have adopted the distance and extinction values from the literature (see Table \ref{table_SNe}). From this comparison, we note that while the brightness at maximum of Event B of SNhunt151 is similar to those of the other transients in the sample (with the exception of the SN impostor SNhunt248 during its ``2014b'' peak), none of them matches the slowly rising Event A. We also note that the Event B light curve of SNhunt151 is broader than those of the other transients, with both a longer rise time and a slower decline. 
 
\begin{figure}
\centering
\includegraphics[width=1\columnwidth]{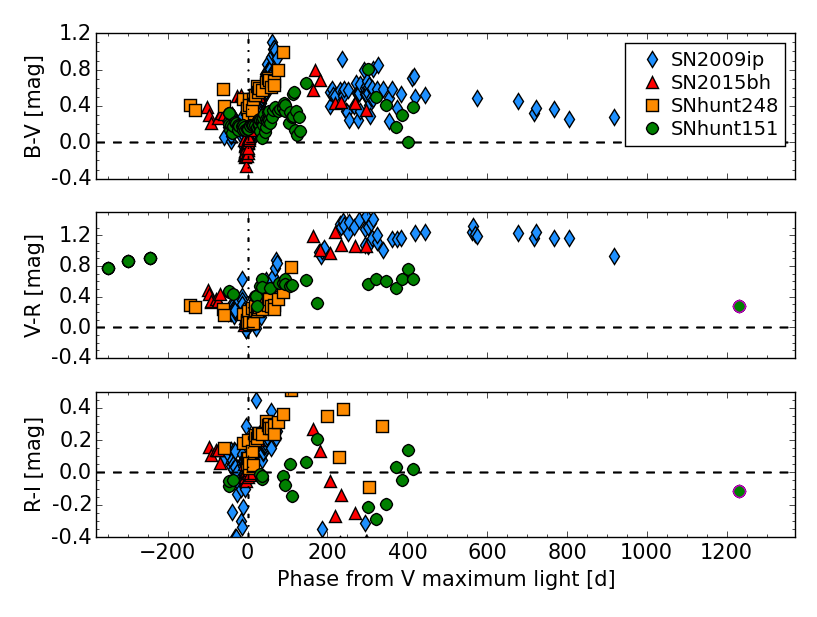}
\caption{Intrinsic colour curves of SNhunt151, compared with those of SN~2009ip, SN~2015bh, and SNhunt248. The dot-dashed vertical line indicates the $V$-band maximum light of SNhunt151. Note that the last epoch of SNhunt151 is an approximation from Sloan magnitudes. A colour version of this figure can be found in the online journal. }
\label{fig_color}
\end{figure}

We also find differences in the colour curves of the objects of our sample (Fig. \ref{fig_color}). SNhunt151 shows a relatively flat and slow colour evolution toward the red for the $(B-V)_0$ and $(V-R)_0$ colours, but an almost constant $(R-I)_0$ colour. The approximated Johnson-Cousins colour at our last epoch (phase 1229 d, adopting the conversions of \cite{chonis08}) also follows this trend. No other object from our sample of comparisons exhibits similar behaviour.

\begin{figure}
\centering
\includegraphics[width=1.\columnwidth]{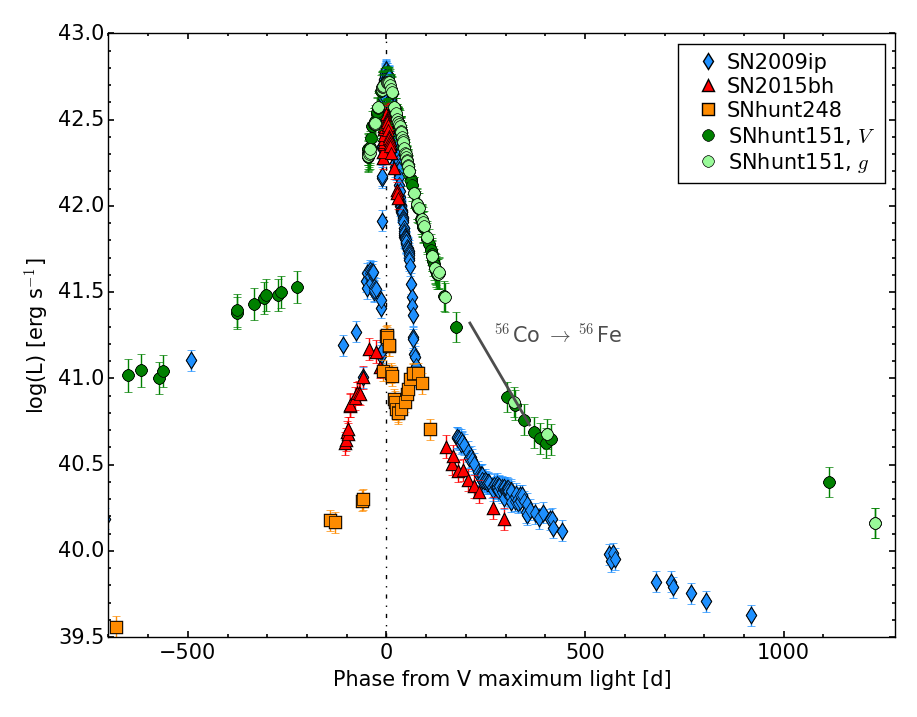}
\caption{Pseudobolometric light curve of SNhunt151 obtained by integrating optical and NIR bands at epochs where $V$ or $g$ were available, compared with those of SN~2009ip, SN~2015bh, and SNhunt248. The dot-dashed vertical line indicates the $V$-band maximum light of SNhunt151. A colour version of this figure can be found in the online journal.}
\label{fig_bol}
\end{figure}

We computed a pseudobolometric light curve of SNhunt151, integrating the wavelength range covered from the $u$ to the $K$ bands (Fig. \ref{fig_bol}). The fluxes at the effective wavelengths were derived from extinction-corrected apparent magnitudes, when observations in the $V$ or $g$ bands were available. When no observation in individual filters was available, the missing photometric epoch was recovered by interpolating the values from epochs close in time or, when necessary, extrapolating assuming a constant colour from the closest available epoch. The pseudobolometric fluxes were estimated integrating the spectral energy distribution (SED) at each epoch following the trapezoidal rule, and assuming zero contribution outside the integration boundaries. Finally, the luminosity was derived from the integrated flux accounting for the adopted distance. The bolometric luminosity errors account for the uncertainties in the distance estimate, the extinction, and the apparent magnitudes. The pseudobolometric light curves of SN~2009ip, SN~2015bh, and SNhunt248 were computed in a similar manner as that of SNhunt151. 

Fitting low-order polynomials to the pseudobolometric light curve of SNhunt151, we estimate a peak luminosity for Event B of $(5.2 \pm 1.3) \times\ 10^{42}$ erg s$^{-1}$. The pseudobolometric peak of SNhunt151 is comparable to that of SN~2009ip, and is more luminous than those of the other transients of the sample (see Fig. \ref{fig_bol} and Table \ref{table_lum}). Note that the Event A maximum is around 15 times fainter than the light-curve peak of Event B. The pseudobolometric tail of SNhunt151 seems to follow roughly the radioactive decay tail of $^{56}$Co to $^{56}$Fe seen in SNe~II, at least until $\sim 400$ d. However, the last two points taken at phase $> 1100$ d are brighter by $\sim 6$ mag than the expected luminosity of a $^{56}$Co-decay-powered light curve.

\begin{table*}
 \centering
  \caption{Peak epochs, peak apparent magnitudes, and decline rates of SNhunt151 in different bands.}
  \label{table_max}
  {
  \begin{tabular}{@{}lccccc@{}}
     \hline
     Band$^a$ & Early rise (Event A) & MJD$_\mathrm{max}$ & $m_\mathrm{max}$ & Decline from max.$^b$ & Tail rate at $\gtrsim 300$ d\\
     &  [mag (100~d)$^{-1}$] & & (mag) & [mag (100~d)$^{-1}$] & [mag (100~d)$^{-1}$] \\
     \hline
    $B$ & - & 56572.0 (0.8) & 18.2 (0.1) & 2.23 (0.26) & 0.21 (0.15) \\
    $V$ & 0.35 (0.15) & 56571.8 (1.0) & 17.4 (0.1) & 2.09 (0.18) & 0.69 (0.11) \\
    $R$ & 0.35 (0.23) & 56572.1 (1.2) & 16.9 (0.1) & 1.92 (0.24) & 0.49 (0.15) \\
    $I$ &   - &              -                      &                  -                      & 1.81 (0.17) & 0.28 (0.08) \\
    \hline
    $u$ &  - & 56568.3 (1.2) & 18.8 (0.1) & 2.39 (0.40) & - \\
    $g$ &  - & 56571.1 (0.4) & 17.8 (0.1) & 2.29 (0.18) & 0.45 (0.23)$^c$\\
    $r$ &  - & 56572.6 (0.8) & 16.9 (0.1) & 1.98 (0.16) & 0.97 (0.19)\\
    $i$ &  - & 56571.9 (0.5) & 16.7 (0.1) & 2.04 (0.14) & 1.32 (0.45)\\
    $z$ & - &      -                      &                  -                      & 2.81 (0.15) & -\\
    \hline
    $J$ &      - &             -                      &                  -                      &  1.85 (0.26) & - \\ 
    $H$ &    - &               -                      &                  -                      & 1.80 (0.23) & - \\ 
    $K$ &     - &              -                      &                  -                      & 1.84 (0.33) & - \\ 
    \hline
  \end{tabular}
  }
  \begin{flushleft} 
  $^a$ Apparent magnitude at peak of the $IzJHK$ light curves could not be constrained. \\
  $^b$ Considering the interval from $\sim 45$ d to 175 d after $V$-band maximum light.\\
  $^c$ Note that the tail rates for the Sloan bands were derived from just two epochs.\\
  \end{flushleft}
\end{table*}

\begin{table}
 \centering
  \caption{Pseudobolometric maximum luminosities.}
  \label{table_lum}
  {
 \begin{tabular}{@{}lc@{}}
\hline
 Object & Luminosity \\
 & ($10^{42}$ erg s$^{-1}$)  \\
\hline
SN~2009ip &  6.1 (1.3) \\
SN~2015bh & 3.3 (0.7) \\
SNhunt248 &   0.2 (0.1) \\
SNhunt151 &   5.2 (1.3) \\
\hline 
\end{tabular}}
\end{table}

%

\section{Spectroscopy}\label{SNspec}

Spectroscopic monitoring of SNhunt151 started soon after the confirmation of the object, on 2013 August 20, and lasted 15 months. We collected a total of 16 optical and 4 NIR spectra. Basic information on the spectroscopic observations can be found in Table \ref{table_spec}. 

All spectra were reduced following standard procedures with {\sc iraf} routines. The two-dimensional frames were debiased and flat-field corrected, before the extraction of the one-dimensional spectra. 

Several pairs of NIR spectra were taken at each epoch at different positions along the slit, and consecutive pairs were subtracted from each other in order to remove the sky background. The subtracted images were aligned to match the spectrum profile and added together. Finally, the source spectrum was extracted from the combined images. 

The one-dimensional optical and NIR spectra were then wavelength calibrated by comparison with arc-lamp spectra obtained during the same night and with the same instrumental configuration, and flux calibrated using spectrophotometric standard stars. The wavelength calibration was verified against the bright night-sky emission lines and adjusted if necessary. The absolute flux calibration of the spectra was cross-checked against the broadband photometry and corrected, when necessary. Finally, the strongest telluric absorption bands in the optical spectra were removed using standard-star spectra (in some cases, residuals are still present after the correction). 

%
\subsection{Spectral Evolution}\label{SNspec_evol}

\begin{figure*}
\centering
\includegraphics[width=1.9\columnwidth]{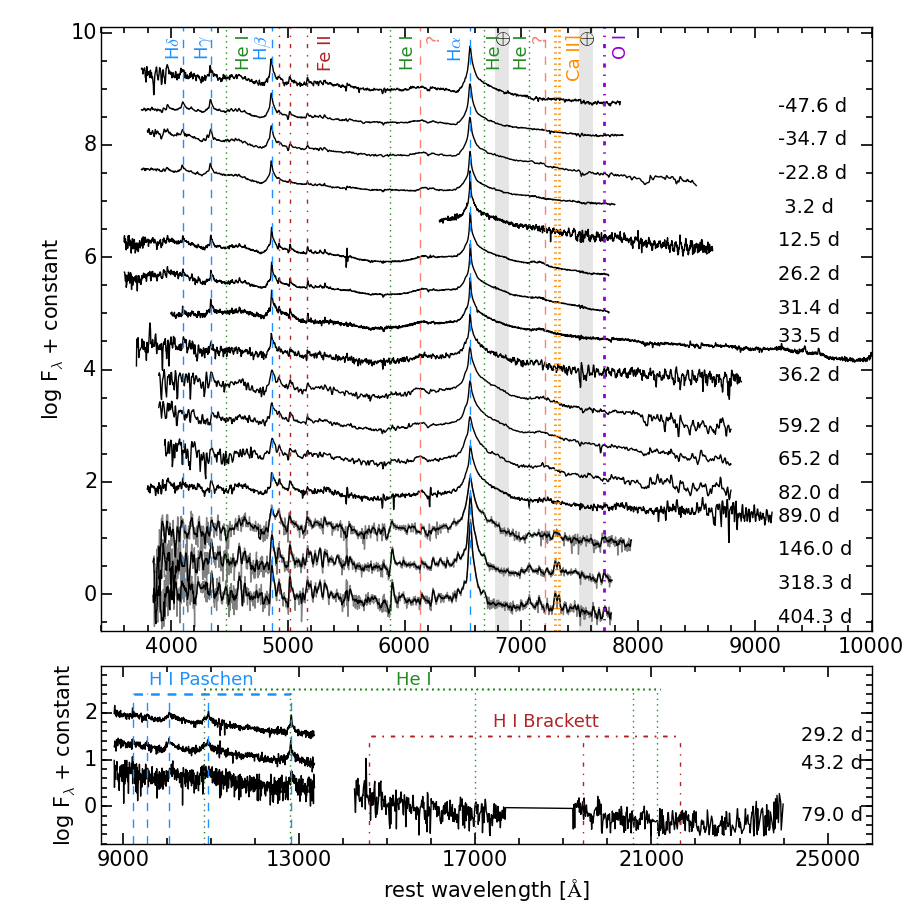}
 \caption{Sequence of optical and NIR spectra of SNhunt151 taken from 2013 August 20 to 2014 November 14. The late-time spectra at 146.0, 318.3, and 404.3 d are shown in grey, with a boxcar-smoothed (using an 8-pixel window) version of the spectra overplotted in black. All spectra have been redshift and extinction corrected. The locations of the most prominent spectral features are indicated by vertical lines. A colour version of this figure can be found in the online journal.}
\label{fig_optnirspec}
\end{figure*}

\begin{figure}
\centering
\includegraphics[width=1.\columnwidth]{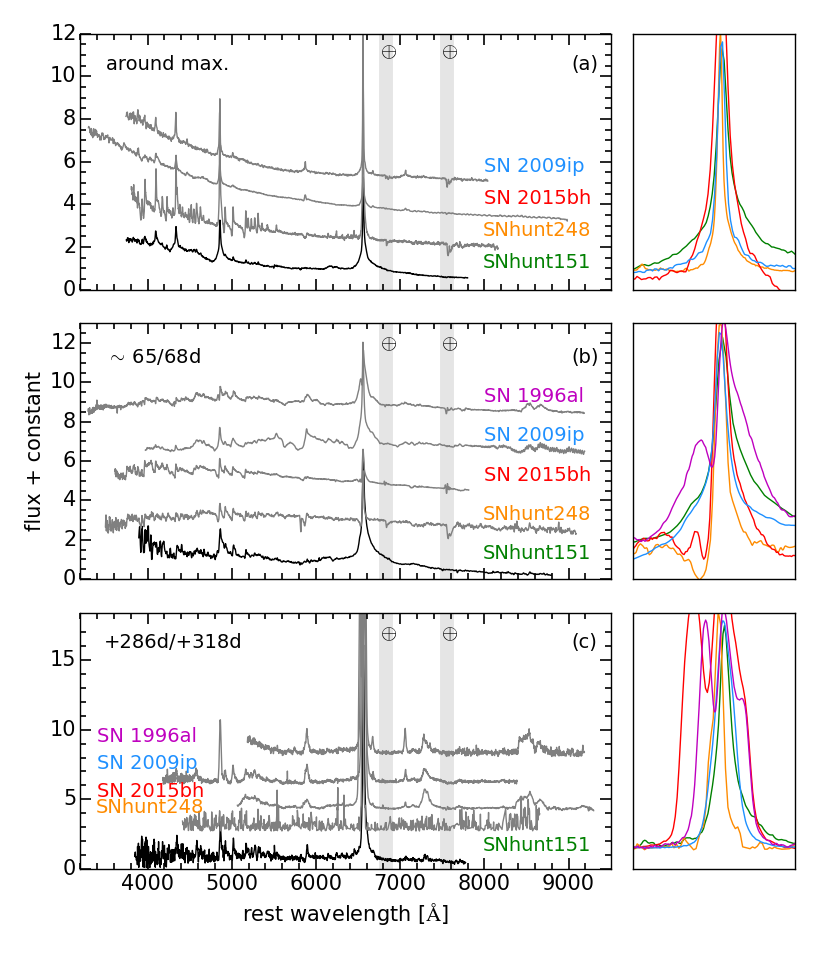}
\caption{Comparison of SNhunt151 spectra around {\it (a)} the maximum of the major peak, and {\it (b)} 65--68 d and {\it (c)} 146--193 d after the peak, with those of the transients SNe~1996al, 2009ip, 2015bh, and SNhunt248 at similar epochs. The H${\alpha}$ profiles are enlarged in the right side of each panel, and normalized to the peak. All spectra have been corrected for their host-galaxy recession velocities and for extinction (values adopted from the literature). A colour version of this figure can be found in the online journal.}
\label{fig_speccomp}
\end{figure}

Figure \ref{fig_optnirspec} shows the sequence of optical and NIR spectra of SNhunt151, while in Figure \ref{fig_speccomp} we compare the optical spectra at representative epochs with those of the Type IIn SN~1996al \citep{benetti16}, and the transients SN~2009ip \citep{pastorello13,fraser13,fraser15}, SN~2015bh \citep{eliasrosa16}, and SNhunt248\footnote{Note that the SNhunt248 phases are considered with respect to its major peak, ``2014b.''} \citep{kankare15} at similar epochs. All of the spectra have been corrected for extinction and redshift using values from the literature (see also Table \ref{table_SNe}).

The spectra of SNhunt151 taken from $\sim -47$ d to $89$ d during Event B exhibit a rather blue continuum and very little evolution. They are dominated by multicomponent Balmer lines in emission, and some relatively weak Fe\,{\sc ii} features. While the Balmer lines do not show evolution during this period, the iron lines become slightly stronger after $65$ d from maximum light. We can also distinguish two broad emission features centered at 6130 $\AA$ and 7200 $\AA$ (they are more evident in the spectra between phases 59 and 82 d), with FWHM $\approx 100$ and 175 $\AA$ ($\sim 4700$ and 8000 \kms), respectively. Similar structure is not seen in H$\beta$. A possible explanation is the presence of a bipolar jet interacting with the dense CSM. This mechanism was also proposed for SN~2010jp, which shows a roughly similar (but rapidly evolving) H$\alpha$ profile \citep{smith12b}.

\begin{figure}
\centering
\includegraphics[width=1.\columnwidth]{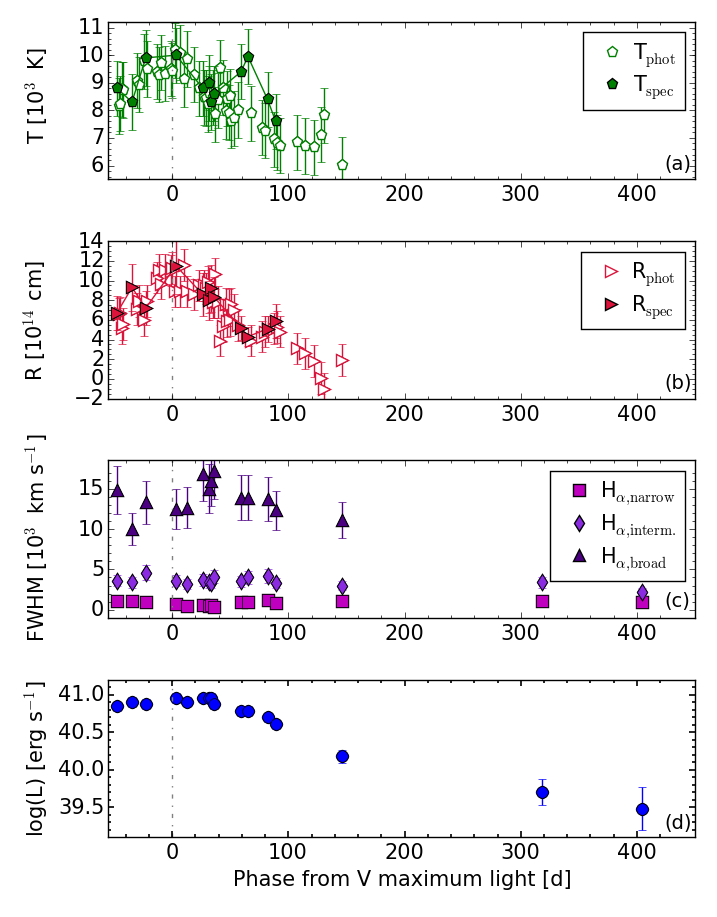}
\caption{{\it (a)}: Evolution of the best-fit blackbody temperatures. {\it (b)}: Evolution of the radius of the photosphere. The solid line connects the temperature and radius measurements. {\it (c)}: FWHM evolution for the broad and narrow H${\alpha}$ emission. {\it (d)}: Evolution of the total luminosity of H${\alpha}$. The dot-dashed vertical line indicates the $V$-band maximum light of SNhunt151. A conservative uncertainty of about $\pm 1000$ K for the temperature has been assumed.}
\label{fig_spectemp_vel}
\end{figure}

The photospheric temperatures at these epochs are estimated by fitting the SED of SNhunt151 with a blackbody function after removing the strongest features of the spectra. As shown in Figure \ref{fig_spectemp_vel}{\it (a)} and in Table \ref{table_specpar}, there is no clear evidence of variation of temperature around the peak. This is in the range $8300$ to $9900$ K (a lower temperature was estimated only for the last spectrum taken just before the seasonal gap, 89~d after maximum light). This behaviour is not seen in other Type IIn SNe (e.g., \citealt{taddia13}), or for the reference transients used here. In all these cases, the temperature is close to 20,000 K at the major peak, decreasing later to 8000--5000~K, and then remaining nearly constant in the following days; see, for example, SN~2009ip \citep{margutti14} and SN~2015bh \citep{eliasrosa16}. We also estimated and plotted the photospheric temperatures derived from the SED of SNhunt151 obtained from broadband photometry. Here there is some evidence of evolution, with a possible slow decrease of temperature after Event B. Similarly, the radius of the photosphere exhibits a likely slow evolution peaking at $11.6 \times\ 10^{14}$ cm during Event B, and then decreases to $1.8 \times\ 10^{14}$ cm at 146~d (Fig. \ref{fig_spectemp_vel}{\it (b))}. At later phases (60 and 80~d after the peak of Event B), the estimated radius seems to increase and then decrease again. This fluctuation of the value of the photospheric radius is possibly caused by a geometric effect (e.g., an asymmetry in the CSM distribution) or a nonlinear CSM density profile (see, e.g., \citealt{benetti16}). Note that these estimates of the radius are approximate, owing to the assumptions made to derive the temperatures and the luminosities of SNhunt151. For instance, the derived temperatures are more likely lower-limit estimates since we have assumed black-body spectra without taking into account effects such as the metal line blanketing.

Around maximum light, SNhunt151 spectra are dominated by Balmer features such as those of the transients SN~2009ip and SN~2015bh. SNhunt151, unlike these transients, does not show He\,{\sc i} features, and its H$\alpha$ profile exhibits a broader component  (Fig. \ref{fig_speccomp}{\it (a)}). At phase 65--68 d (Fig. \ref{fig_speccomp}{\it (b)}), SNhunt151 is still relatively blue, while the other transients at the same phase turned red. Another difference to be noted is the absence of P-Cygni minima in the iron and Balmer lines.

At late phases ($> 145$ d), the broader component of the H$\alpha$ line seems to be weakening, while the spectra are still dominated by the narrow component (see Section \ref{SNspec_evolhalpha}). Resolved narrow lines ($\sim 900$ \kms\ FWHM) of He\,{\sc i} $\lambda$5876, $\lambda$6678, and $\lambda$7065 are now evident in the spectra, as well as [Ca\,{\sc ii}] $\lambda\lambda$7291, 7323. A hint of O\,{\sc i} $\lambda$7774 might be identified, but because of the low S/N in that region, it is only a tentative identification. Fe\,{\sc ii} lines are still strong and show low-contrast P-Cygni profiles (for example, Fe\,{\sc ii} $\lambda$4923, $\lambda$5018, and $\lambda$5169) with expansion velocities between 900 and 1400 \kms\ from the absorption minima.

These features are still weak if we compare SNhunt151 spectra with those of the transients of our sample at similar epochs (Fig. \ref{fig_speccomp}{\it (c)}). We also highlight the differences between the H${\alpha}$ profiles (see inset in Fig. \ref{fig_speccomp}{\it (c)}), with SNhunt151 showing a quite symmetric profile. This is similar to that of SN~2009ip, except for an extended red wing of SNhunt151, and unlike the asymmetric H${\alpha}$ profiles of SNhunt248 and SN~2015bh, or the double-peaked one of SN~1996al.

The NIR spectra of SNhunt151 (Fig. \ref{fig_optnirspec}, lower panel) are dominated by the hydrogen Paschen series. A trace of Br$\gamma$ and He\,{\sc i} features may also be detected. In fact, comparing the profile of the strongest Paschen lines of our spectrum at phase $\sim 30$ d with H$\alpha$ at a similar epoch (Fig. \ref{fig_spechalpha}{\it (e)}), we can see how the Paschen line at $\lambda$10940 is clearly blended with He\,{\sc i} $\lambda$10,830, while the line at 12,820~\AA\ is similar to H$\alpha$ but apparently without the broad red wing. As at optical wavelengths, no P-Cygni profiles are seen.


\begin{table*}
 \centering
  \caption{Main parameters as inferred from the spectra of SNhunt151. The velocities are computed from the decomposition of the H${\alpha}$ profile.}
  \label{table_specpar}
  \scalebox{0.85}{
 \begin{tabular}{@{}ccrcccccc@{}}
\hline
 Date & MJD & Phase & Temperature$^a$ & FWHM$_\mathrm{H\alpha,nar}^b$ & FWHM$_\mathrm{H\alpha,inter}$ & FWHM$_\mathrm{H\alpha,br}$ & Luminosity$_\mathrm{H\alpha}$ & EW$_\mathrm{H\alpha}$$^c$ \\
 & & (days) & (K)  & (\kms) &  (\kms) & (\kms) & ($10^{39}$ erg s$^{-1}$) & $\AA$\\
\hline
20130820 & 56524.20 & $-$47.6 & 8800 & 1020 (204)  & 3600 (720)  & 14,860 (2972) & 70 (10) & 300 (60) \\
20130902 & 56537.18 & $-$34.7 & 8300 & 1070 (214)  & 3400 (680)  & 10,000 (2000) & 80 (12) & 240 (50) \\
20130914 & 56549.01 & $-$22.8 & 9900 &  970 (194)  & 4600 (920)  & 13,300 (2660) & 75 (11) & 195 (40) \\
20131010 & 56575.07 & 3.2    & 8600 &  650 (130)  & 3600 (720)  & 12,500 (2500) & 90 (13) & 190 (40) \\
20131019 & 56584.31 & 12.5   &  -   &  430 (86)   & 3150 (630)  & 12,650 (2530) & 80 (11) & 200 (40) \\
20131102 & 56598.05 & 26.2   & 8800 &  580 (116)  & 3700 (740)  & 16,800 (3360) & 90 (12) & 200 (40) \\
20131107 & 56603.26 & 31.4   & 9000 &  450 (90)   & 3400 (680)  & 15,000 (3000) & 90 (12) & 220 (45) \\
20131109 & 56605.39 & 33.5   & 8300 &  570 (114)  & 3300 (660)  & 16,000 (3200) & 90 (13) & 230 (45) \\
20131112 & 56608.04 & 36.2   & 8600 &  390 (78)   & 4100 (820)  & 17,200 (3440) & 75 (11) & 260 (50) \\
20131205 & 56631.07 & 59.2   & 9400 &  910 (182)  & 3500 (700)  & 13,900 (2780) & 60 (9) & 270 (50) \\ 
20131211 & 56637.00 & 65.2   & 9950 &  980 (196)  & 4000 (800)  & 13,900 (2780) & 60 (9) & 290 (60) \\ 
20131227 & 56653.85 & 82.0   & 8400 & 1220 (244)  & 4200 (840)  & 13,700 (2740) & 50 (7) & 280 (55) \\ 
20140103 & 56660.87 & 89.0   & 7600 &  840 (168)  & 3300 (660)  & 12,300 (2460) & 40 (6) & 270 (55) \\ 
20140301 & 56717.88 & 146.0  &   -  & 1070 (214)  & 3000 (600)  & 11,100 (2220) & 15 (3) & 285 (55) \\ 
20140821 & 56890.19 & 318.3  &   -  & 1020 (204)  & 3400 (680)  &     -        &  5 (2) & 280 (55) \\ 
20141115 & 56976.16 & 404.3  &   -  &  900 (180)  & 2200 (440)  &     -        &  3 (2) & 300 (60) \\ 
\hline
\end{tabular}}
\begin{flushleft}
$^a$ We consider a conservative uncertainty in the temperature of about $\pm$ 1000 K.\\
$^b$ We consider a conservative uncertainty in the velocities of about 20\%.\\
$^c$ We consider a conservative uncertainty in the EW of about 20\%.\\
\end{flushleft}
\end{table*}

%

\subsection{Evolution of the Balmer Lines}\label{SNspec_evolhalpha}

\begin{figure*}
\centering
\includegraphics[width=1.9\columnwidth]{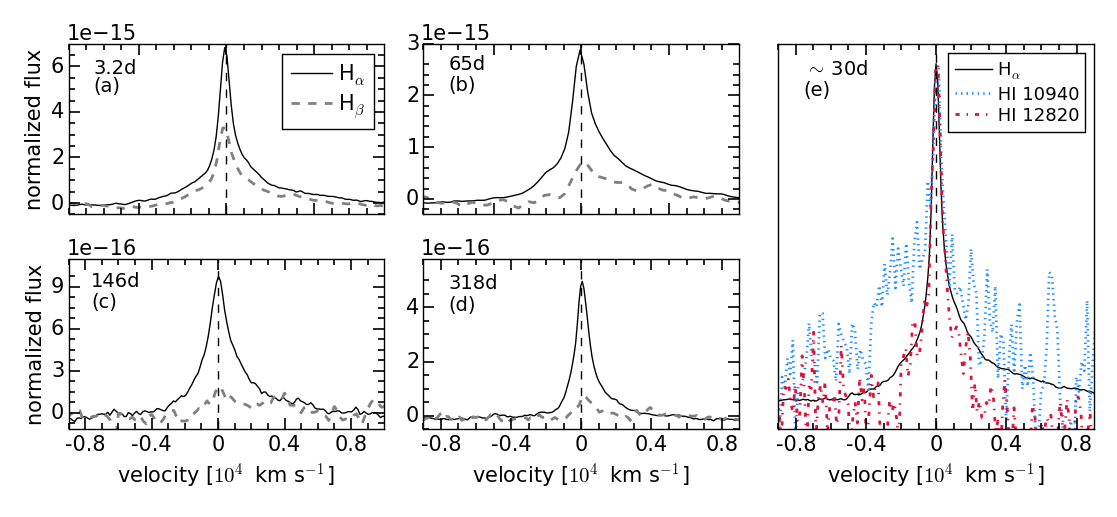}
 \caption{{\it (a--d)}: H$\alpha$ and H$\beta$ line profiles at representative epochs. {\it (e)}: H$\alpha$ and Paschen $\lambda$10,940 and $\lambda$12,820 lines at phase $\sim 30$ d, normalized to the peak. The dashed lines mark the rest wavelength of H${\alpha}$. }
\label{fig_spechalpha}
\end{figure*}

\begin{figure}
\centering
\includegraphics[width=1.\columnwidth]{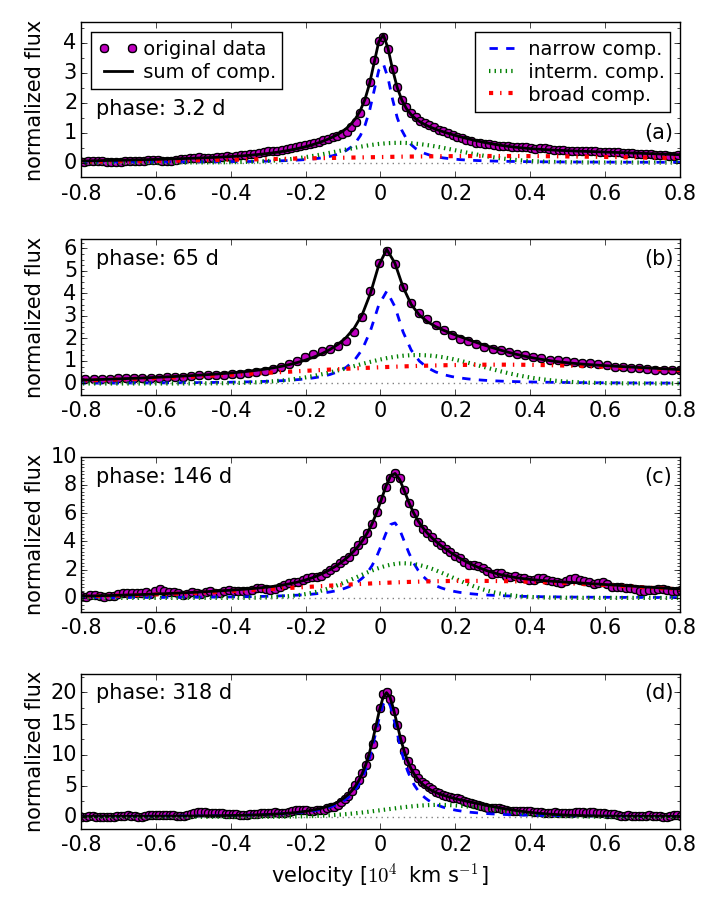}
\caption{Decomposition of the H${\alpha}$ emission line of SNhunt151 at 3.2, 65, 146, and 318 d from $V$-band maximum light. A colour version of this figure can be found in the online journal.}
\label{fig_specdecomp}
\end{figure}

H$\alpha$ is the most prominent line in the SNhunt151 spectra, and the analysis of its profile may allow us to probe the environment of the transient. As we can see in Figure \ref{fig_spechalpha}, the H$\alpha$ profile seems to consist of multiple components which do not show relative evolution with time. H$\beta$ exhibits a similar profile at earlier phases, but it is probably contaminated by iron lines at phases later than $\sim 50$ d. Following a procedure described by \citealt{eliasrosa16} for SN~2015bh, we decompose the line profile at all epochs into three emission components. We used a Lorentzian profile for the narrow component, and Gaussian functions for the intermediate and broad components, except for the two latest-time spectra in which there is no evidence for a broad component. Figure \ref{fig_specdecomp} displays the results of the multicomponent fit at some representative epochs: 3.2 d, 65 d, 146 d, and 318 d. The derived velocities are listed in Table \ref{table_specpar}, and their evolution is shown in Figure \ref{fig_spectemp_vel}{\it (c)}. 

The FWHM of the narrow H$\alpha$ emission\footnote{The narrow-line component was resolved in all spectra. We first corrected the measured FWHM for the spectral resolution, and then computed the velocity.} remains nearly constant, with a velocity of $\sim 500$ \kms\ as determined from our highest-resolution spectra ($\sim 150$--215~\kms). The intermediate component, related to the shocked ejecta region between the forward shock and the reverse shock, also remains roughly constant with an average FWHM of $\sim 3600$ \kms. The broader component also exhibits an apparently constant evolution ($\sim 14,000$ \kms\ FWHM), and then disappears in the last two spectra ($> 318$ d). 

The velocity of the broad component of H$\alpha$ is high in comparison with the velocities of the H-rich material expelled in major eruptions of massive stars such as luminous blue variables (LBVs; e.g., \citealt{smith08,pastorello13}). It is more indicative of a SN explosion, but in this case the lack of evolution is rather puzzling. In fact, it is likely possible that the broad component is the Lorentzian wing of the narrow line, owing to electron scattering in the opaque CSM (e.g., \citealt{chugai01}). 

Interestingly, the total luminosity of the H$\alpha$ line (Fig. \ref{fig_spectemp_vel}{\it (d)}, and Table \ref{table_specpar}) remains almost constant at $\sim 80 \times 10^{39}$ erg s$^{-1}$ until $\sim 50$ d, and decreases thereafter to $\sim 3 \times 10^{39}$ erg s$^{-1}$ in our last spectrum.

%

\section{Discussion and Summary}\label{SNdiscus}

In the previous sections, we analysed the observed properties of SNhunt151. This transient shows a slow rise of around 450 d, ranging from a  $V$-band absolute magnitude of approximately $-13.8$ mag to $-15.0$ mag, then followed by a major rebrightening which peaks at $-18.1$ mag. Recent {\sl HST}+WFC3 and GTC+OSIRIS images, taken around 3~yr after maximum, show a source at the position of SNhunt151 of $\sim -12$ mag in $F555W$ ($\sim V$) and $-11.6$ mag in $g$. These late epochs do not follow the typical radioactive heating expected in SNe~II, but instead indicate a shallower decline. During Event B, the spectra are always dominated by Balmer lines, and show negligible evolution.

As for other similar transients found in the last decade (e.g., SN~2009ip), it is arduous finding a unique interpretation for the chain of events of SNhunt151. As seen in previous sections, some observables of SNhunt151 could resemble those of the SN~2009ip-like transients. However, there are also significant differences, as follows.

\begin{enumerate}
\item[(1)] No significant variability brighter than $-13$ mag in the historical light curve (from the end of 2003 to the end of 2011). 
\item[(2)] Slow prediscovery rise (Event A).
\item[(3)] Broad Event B with respect to other SN~2009ip-like transients.
\item[(4)] Stronger ejecta-CSM interaction (or a thicker envelope) over its entire evolution.
\item[(5)] No clear evidence of evolution in the temperature. 
\item[(6)] No evidence of P-Cygni profiles in the Balmer lines during the entire evolution, and in iron features until phase $\sim 90$ d.
\end{enumerate}

A possible scenario is that during the outbursts, SNhunt151 is surrounded by a nebula or cocoon of material hiding the inner engine. In some respects, this resembles the case of the Homunculus Nebula around $\eta$ Carinae (e.g., \citealt{smith12}). The nebula would enclose a recent event that can be almost anything, such as a collision between massive shells ejected by a very massive star, a merger event in a close binary system, or a terminal core-collapse SN explosion (either during Event A with the successive strong interaction of the SN ejecta and a circumstellar shell, or during Event B preceded by a giant eruption in Event A). One of the possible explanations for the lack of spectral evolution is that we are observing an asymmetric environment from an unfavorable orientation, where the more dense material conceals what is below.

Using the pseudobolometric luminosity, we attempt to constrain the total energy radiated by SNhunt151. We estimate an energy of $7.4 \times 10^{48}$ erg during  Event A, and $4.2 \times 10^{49}$ erg during  Event B. The latter radiated energy is comparable to that of SNe~IIn such as SN~2011ht ($\sim 3 \times 10^{49}$ erg; \citealt{roming12}), or even the luminous SN~1998S ($\sim 9 \times 10^{49}$ erg; \citealt{liu00,fassia00,pozzo04}). These values are also similar to the high energy of other types of transients, such as the controversial SN~2009ip ($\sim 1.5 \times 10^{48}$ erg and $\sim 3 \times 10^{49}$ erg for its respective events 2012a and 2012b; \citealt{margutti14}), or SN~1961V ($\sim 4 \times 10^{48}$ erg during the  prepeak plateau, and $\sim 6 \times 10^{49}$ erg during the major and brief peak; \citealt{bertola63,bertola64,bertola65,bertola67}). The ``Great Eruption" of $\eta$ Carinae released a radiated energy of $\sim 5 \times 10^{49}$ erg (e.g., \citealt{humphreys12p}); however, it is noteworthy that the eruption lasted 20~yr. Summarising, the amount of energy radiated during both Event A and Event B of SNhunt151 is not a conclusive argument to assess whether the object is a terminal event.

The main puzzle is Event A. Slow rising light curves have been previously seen in some Type IIn SNe, such as SN~2008iy \citep{miller10}, but with a luminosity at the peak which is a factor of 2.5 brighter than the maximum  luminosity of  SNhunt151 Event A. Given the SNhunt151 observables, it seems unavoidable that there is a considerable amount of opaque material between the engine and the observer. Thus, the diffusion of the photons through the material is slow. Following the relations reported by \citealt{ofek13b}, from the rise time of the possible SN explosion (i.e., from Event B of SNhunt151, $\sim 50$~d) we can estimate an upper limit to the amount of mass between the transient and us. Assuming as wind velocity the value inferred for the narrow H$\alpha$  component during that rise ($\sim 1000$ km s$^{-1}$), we derive a mass-loss rate $< 1.1$ M$_{\sun}$ yr$^{-1}$. This value is comparable to that of LBV giant eruptions (e.g., \citealt{smith16c}).

\begin{figure}
\centering
\includegraphics[width=1\columnwidth]{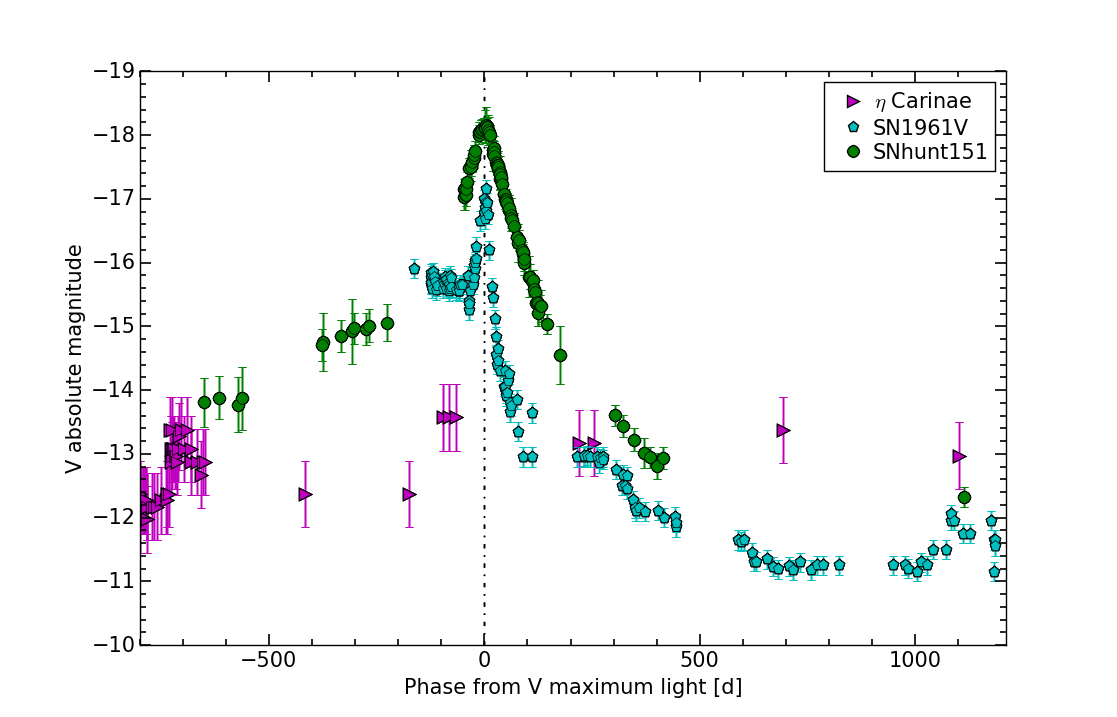}
\caption{Absolute $V$ light curve of SNhunt151 ({\it circles}), shown along with those of supernova (or impostor) SN~1961V ({\it pentagons}; photographic mag) and the revised visual light curve of the Great Eruption of $\eta$ Carinae ({\it rotated triangles}). The dot-dashed vertical line indicates the $V$-band maximum light of SNhunt151. A colour version of this figure can be found in the online journal.}
\label{fig_abs61V}
\end{figure}

One interesting comparison is with SN~1961V. As we see in Figure \ref{fig_abs61V}, the magnitude at the end of  Event A of SNhunt151 is suggestively similar to that of the pseudoplateau shown by SN~1961V before its peak. The origin of this first event in SN~1961V depends on the interpretation. If SN~1961V was a true core-collapse SN \citep{smith11d,kochanek11}, then the light curve could be a combination of a Type II-Plateau-like SN, followed by a peak arising from interaction of the SN ejecta with the CSM. Instead, if the SN~1961V is a SN impostor (e.g., \citealt{vandyk_rev12,vandyk12}), the pre-peak event is due to a series of minor eruptions and interaction with the material previously ejected. 

An argument against the interpretation of SN~1961V as a SN impostor is the lack of detectable mid-infrared (MIR) emission at the transient position, expected from the dust shell formed around surviving LBVs \citep{kochanek11}. A counterexample is $\eta$~Carinae, which is extremely luminous at IR wavelengths (e.g., \citealt{smith12}). 

Looking for some evidence of dust around SNhunt151, we inspected the {\sl Spitzer Space Telescope} Heritage Archive\footnote{\url{http://sha.ipac.caltech.edu/applications/Spitzer/SHA/.}} after the transient discovery. We found images at four epochs: 2005 February 23 and 24 (3.6, 4.5, 5.8, and 8.0 $\mu$m channels; Programme ID 3584, PI D.~Padgett), 2014 November 27 (3.6 and 4.5 $\mu$m channels; Programme ID 10139, PI O.~D.~Fox), and 2015 May 23 (3.6 and 4.5 $\mu$m channels; Programme ID 11053, PI O.~D.~Fox), taken with the InfraRed Array Camera (IRAC) aboard {\sl Spitzer}. We worked with the {\it Post Basic Calibrated Data (pbcd)}, which are already fully coadded and calibrated. We did not detect a source at the SNhunt151 position in the first two epochs (year 2005); however, a source is well detected in the last two (years 2014 and 2015) in both {\sl Spitzer} channels. We estimate the integrated flux of the source (in the observations of 2005 we use an aperture of $3 \times 3$ pixels around the transient position) using  {\sc MOPEX}\footnote{{\sc MOPEX} is a {\sl Spitzer} software package: \url{http://irsa.ipac.caltech.edu/data/SPITZER/docs/dataanalysistools/tools/mopex/.}}. Following the ``recipe'' advised by the {\sl Spitzer} team, we obtained the values reported in Table \ref{table_spitzer}. The late-time NIR detections may suggest the formation of dust in a cool dense shell in SNhunt151, as in $\eta$ Carinae and other SNe~IIn (e.g., \citealt{fox11,fox13}) or, alternatively, an IR echo produced by foreground, pre-existing dust.

In summary, here we have discussed our observations of a unique object, SNhunt151. Its nature is still uncertain, but we support the hypothesis that we are observing a dense nebula of material and dust at the position of the transient, which reprocesses radiation from the underlying energetic outburst and enshrouds the mechanism that gives rise to the chain of events. It is clear that we should still keep an eye on SNhunt151 in the future.


\begin{table*}
\caption{{\sl Spitzer} integrated fluxes of the SNhunt151 field.}
\label{table_spitzer}
\scalebox{0.85}{
\begin{tabular}{@{}cccccccc@{}}
\hline 
Date & MJD & Phase$^a$ & 3.6 $\mu$m & 4.5 $\mu$m & 5.8 $\mu$m & 8.0 $\mu$m & Program ID/P.I.\\
 &  & (days) & ($\mu$Jy) & ($\mu$Jy) & ($\mu$Jy) & ($\mu$Jy) & \\
\hline  
20050223 & 53424.87 & $-3147.0$ & $< 49.8$ & $< 29.0$ & $< 567.8$ & $< 1290.0$ & 3584/D.~Padgett\\ 
20050224 & 53425.39 & $-3146.5$ & $< 63.4$ & $< 46.0$ & $< 699.6$ & $< 1298.0$ & 3584/D.~Padgett\\ 
20141127 & 56988.00 & 416.2 & $14.5 (0.2)$ & $20.7 (0.3)$ & - & - & 10139/O.~D.~Fox\\
20150523 & 57165.81 & 594.0 & $12.6 (0.4)$ & $14.6 (0.3)$ & - & - & 11053/O.~D.~Fox \\
\hline 
\end{tabular}}
\begin{flushleft}
$^a$ Phases are relative to $V$ maximum light, MJD $= 56571.8 \pm 1.0$.\\ 
\end{flushleft}
\end{table*}

%
\section*{Acknowledgements}

We thank S. Spiro, R. Rekola, A. Harutyunyan, and M. L. Graham for their help with the observations. We are grateful to the collaboration of Massimo Conti, Giacomo Guerrini, Paolo Rosi, and Luz Marina Tinjaca Ramirez from the Osservatorio Astronomico Provinciale di Montarrenti. The staffs at the different observatories provided excellent assistance with the observations.

The research leading to these results has received funding from the European Union Seventh Framework Programme (FP7/2007-2013) under grant agreement No. 267251, ``Astronomy Fellowships in Italy'' (AstroFIt).'' N.E.-R. acknowledges financial support from MIUR PRIN 2010-2011, ``The Dark Universe and the Cosmic Evolution of Baryons: From Current Surveys to Euclid." N.E.-R., A.P., S.B., L.T., M.T., and G.T are partially supported by the PRIN-INAF 2014 (project ``Transient Universe: Unveiling New Types of Stellar Explosions with PESSTO''). G.P. acknowledges support provided by the Millennium Institute of Astrophysics (MAS) through grant IC120009 of the Programa Iniciativa Cient\'{i}ifica Milenio del Ministerio de Econom\'{i}a, Fomento y Turismo de Chile. T.K. acknowledges financial support from the Emil Aaltonen Foundation. CRTS was supported by the NSF grants AST-0909182, AST-1313422, and AST-1413600. A.V.F. is grateful for generous financial assistance from the Christopher R. Redlich Fund, the TABASGO Foundation, the Miller Institute for Basic Research in Science (UC Berkeley), and NASA/HST grant GO-14668 from the Space Telescope Science Institute, which is operated by AURA, Inc. under NASA contract NAS5-26555. The work of A.V.F. was conducted in part at the Aspen Center for Physics, which is supported by NSF grant PHY-1607611; he thanks the Center for its hospitality during the neutron stars workshop in June and July 2017. N.E.-R. acknowledges the hospitality of the ``Institut de Ci\`encies de l'Espai" (CSIC), where this work was completed.

This research is based on observations made with the Nordic Optical Telescope, operated by the Nordic Optical Telescope Scientific Association at the Observatorio del Roque de los Muchachos, La Palma, Spain, of the Instituto de Astrof\'{i}sica de Canarias; 
the Gran Telescopio Canarias (GTC), installed in the Spanish Observatorio del Roque de los Muchachos of the Instituto de Astrof\'{i}sica de Canarias, on the island of La Palma;
the Italian Telescopio Nazionale Galileo (TNG), operated on the island of La Palma by the Fundación Galileo Galilei of the INAF (Istituto Nazionale di Astrofisica) at the Spanish Observatorio del Roque de los Muchachos of the Instituto de Astrof\'{i}sica de Canarias;
the Liverpool Telescope, operated on the island of La Palma by Liverpool John Moores University in the Spanish Observatorio del Roque de los Muchachos of the Instituto de Astrof\'{i}sica de Canarias with financial support from the UK Science and Technology Facilities Council; 
the 1.82-m Copernico Telescope and the Schmidt 67/92 cm of INAF-Asiago Observatory;
the Catalina Real Time Survey (CRTS) Catalina Sky Survey (CSS) 0.7 m Schmidt Telescope; and the Las Cumbres Observatory (LCO) network. 
This work is also based in part on archival data obtained with the NASA/ESA {\sl Hubble Space Telescope}, obtained from the Data Archive at the Space Telescope Science Institute (STScI), which is operated by the Association of Universities for Research in Astronomy (AURA), Inc., under NASA contract NAS5-26555; the {\sl Spitzer Space Telescope}, which is operated by the Jet Propulsion Laboratory, California Institute of Technology, under a contract with NASA (support was provided by NASA through an award issued by JPL/Caltech); and the {\sl Swift} telescope.

This work has made use of the NASA/IPAC Extragalactic Database (NED), which is operated by the Jet Propulsion Laboratory, California Institute of Technology, under contract with NASA.




\bibliographystyle{mnras}
\bibliography{impostors}



\appendix

\section[]{Properties of the supernovae used in this work.}

\begin{table*}
 \centering
  \setlength\tabcolsep{2.5pt}
  \caption{Properties of the transients used in this work.}\label{table_SNe}
  \begin{tabular}{@{}llccccc@{}}
  \hline
 Object  & Host Galaxy & Redshift & Distance$^{\dagger}$ & $E(B-V)_{\rm tot}$  &  $V$ max date & Sources  \\
        &  & & (Mpc) & (mag) & (MJD) & \\
 \hline
SN~1961V & NGC~1058 & 0.002 & 9.3 & 0.05 & 37641.0 & a \\
 $\eta$ Car & - & 0.00008 & 0.0023 & 0.24 & $\sim -5000.0$ & b \\
SN~1996al & NGC~7689 & 0.007 & 22.9 & 0.11 & 50265.0 & c \\
SN~2009ip & NGC~7259 & 0.006 & 25.0 & 0.02 & 56207.7 &  d \\
SN~2015bh & NGC~2770 & 0.007 & 29.3 & 0.21 & 57166.9 & e \\
SNhunt248 & NGC~5806 & 0.005 & 21.9 & 0.05 & 56828.0$^{\ddagger}$ & f \\
SNhunt151 & UGC~3165 & 0.013 & 51.8 & 0.64 &  56571.8 & This work \\
\hline
\end{tabular}
\begin{flushleft}
$^{\dagger}$ Distances have been scaled to H$_0$ = 73 \kms\ Mpc$^{-1}$.\\
$^{\ddagger}$ This epoch corresponds to the ``2014b'' peak of SNhunt248.\\
a = \citealt{bertola63,bertola64,bertola65,bertola67}; b = \citealt{smith11c,humphreys12p}; c = \citealt{benetti16}; d = \citealt{maza09,smith10,foley11,pastorello13,mauerhan13a,fraser13}; e = \citealt{eliasrosa16}; f=\citealt{kankare15}. 
\end{flushleft}
\end{table*}

%
\section[]{Tables of photometry and spectroscopy of SNhunt151}

\begin{table*}
\centering
\caption{Basic information about the telescopes and instruments used (in alphabetical key order).}
\label{table_setup}
\scalebox{0.7}{
\begin{tabular}{@{}lllll@{}}
\hline
Tables Code & Telescope & Instrument & Pixel Scale & Location \\
 &  &  & (arcsec pixel$^{-1}$) & \\
\hline
AFOSC & 1.82 m Copernico Telescope & AFOSC & 0.52 & Mount Ekar, Asiago, Italy\\
ALFOSC& 2.56 m Nordic Optical Telescope & ALFOSC  & 0.19 &  Roque de Los Muchachos, La Palma, Canary Islands, Spain\\
Apogee & 0.53 m Ritchey-Chretien Telescope & Apogee & 1.17 & Osservatorio Astronomico Provinciale di Montarrenti, Siena, Italy\\
FLI-New & 0.50 m Newtonian Telescope & FLI Proline CCD & 2.32 & Osservatorio di Monte Agliale,Lucca, Italy\\
IO:O & 2.00 m Liverpool Telescope & IO:O   & 0.30 & Roque de Los Muchachos, La Palma, Canary Islands, Spain\\
kb71 & 1.00 m-03 Telescope & kb71 & 0.47 & LCOGT$^a$ node at Siding Spring Observatory, Australia\\
kb77 & 1.00 m-04 Telescope & kb77 & 0.47 & LCOGT node at Cerro Tololo Inter-American Observatory, Chile\\
kb78 & 1.00 m-05 Telescope & kb78 & 0.47 & LCOGT node at Cerro Tololo Inter-American Observatory, Chile\\
kb74 & 1.00 m-08 Telescope & kb74 & 0.47 & LCOGT node at Cerro Tololo Inter-American Observatory, Chile\\
kb73 & 1.00 m-09 Telescope & kb73 & 0.47 & LCOGT node at Cerro Tololo Inter-American Observatory, Chile\\
kb70 & 1.00 m-10 Telescope & kb70 & 0.47 & LCOGT node at South African Astronomical Observatory, South Africa\\
kb05 & 1.00 m-11 Telescope & kb05 & 0.47 & LCOGT node at Siding Spring Observatory, Australia\\
kb75 & 1.00 m-12 Telescope & kb75 & 0.47 & LCOGT node at South African Astronomical Observatory, South Africa\\
kb79 & 1.00 m-13 Telescope & kb79 & 0.47 & LCOGT node at South African Astronomical Observatory, South Africa\\
kb76 & 1.00 m-13 Telescope & kb76 & 0.47 & LCOGT node at South African Astronomical Observatory, South Africa\\
LRS & 3.58 m Telescopio Nazionale Galileo & LRS  & 0.25 &  Roque de Los Muchachos, La Palma, Canary Islands, Spain\\
NOTCam & 2.56 m Nordic Optical Telescope & NOTCam & 0.24 & Roque de Los Muchachos, La Palma, Canary Islands, Spain\\
NICS & 3.58 m Telescopio Nazionale Galileo & NICS & 0.25 &  Roque de Los Muchachos, La Palma, Canary Islands, Spain\\
OSIRIS & 10.4 m Gran Telescopio CANARIAS & OSIRIS  & 0.25 &  Roque de Los Muchachos, La Palma, Canary Islands, Spain\\
Palomar & 5.10 m Hale Telescope & DBSP & 0.39 & Palomar Observatory, California, USA\\
PTF & 1.20 m Samuel Oschin Telescope & CCD & 1.01 & Palomar Observatory, California, USA\\
RATCam & 2.00 m Liverpool Telescope & RATCam & 0.28 & Roque de Los Muchachos, La Palma, Canary Islands, Spain\\
SBIG & 0.67/0.92 m Schmidt Telescope & SBIG & 0.86 & Mount Ekar, Asiago, Italy \\
SDSS & 2.50 m Telescope & Sloan Digital Sky Survey CCD & 0.39 & Apache Point Observatory, New Mexico, USA\\
SI & 1.52 m Cassegrain reflector &  SI 600-386 & 0.96 & Catalina Sky Survey node at Mount Lemmon Observatory, USA\\
SOAR & 4.10 m Southern Astrophysical Research Telescope & Goodman & 0.15 & Cerro Tololo Inter-American Observatory, Chile\\
Spitzer & 0.8-m Spitzer Space Telescipe & IRAC & 0.60 & - \\
T17FLI & 0.43 m iTelescope.Net T17  & FLI-New-Beta & 0.92 & Siding Spring Observatory, Australia\\
T18SBIG & 0.32 m iTelescope.Net T18  & SBIG STL-6303 3 CCD Camera  & 0.63 & AstroCamp Observatory, Nerpio, Spain\\
T21FLI & 0.43 m iTelescope.Net T21  & FLI-New & 0.96 & Mayhill, New Mexico, USA\\
T7SBIG & 0.43 m iTelescope.Net T7   & SBIG STL-11000 3 CCD Camera & 0.73 & AstroCamp Observatory, Nerpio, Spain\\
T9SBIG & 0.32 m iTelescope.Net T9   & SBIG ST-8 3 CCD Camera	 & 0.80 & AstroCamp Observatory, Nerpio, Spain\\
WFC3 & 2.40 m Hubble Space Telescope & WFC3/UVIS & 0.04 & - \\
\hline
\end{tabular}
}
\begin{flushleft}
$^a$ LCOGT = Las Cumbres Observatory Global Telescope Network.\\
\end{flushleft}
\end{table*}

\begin{table*}
\caption{Optical Johnson-Cousins photometry of SNhunt151 ({\sc Vegamag}).}
\label{table_JCph}
\begin{tabular}{@{}cccccccc@{}}
\hline 
Date & MJD & Phase$^a$  & $B$ & $V$ & $R$  & $I$ & Instrument Key \\ 
 &  & (days) & (mag) & (mag) & (mag)  & (mag) &  \\ 
\hline 
20051028 & 53671.48 & -2900.4 & - &  $>  22.5 $&  - &  - & SI  \\ 
20051227 & 53731.40 & -2840.4 & - &  $>  21.8 $&  - &  - & SI  \\ 
20060125 & 53760.44 & -2811.4 & - &  $>  21.4 $&  - &  - & SI  \\ 
20060126 & 53761.23 & -2810.6 & - &  $>  21.4 $&  - &  - & SI  \\ 
20060202 & 53768.29 & -2803.6 & - &  $>  22.1 $&  - &  - & SI  \\ 
20060303 & 53797.32 & -2774.5 & - &  $>  21.6 $&  - &  - & SI  \\ 
20060928 & 54006.43 & -2565.4 & - &  $>  21.6 $&  - &  - & SI  \\ 
20061020 & 54028.43 & -2543.4 & - &  $>  22.0 $&  - &  - & SI  \\ 
20061101 & 54040.31 & -2531.5 & - &  $>  21.8 $&  - &  - & SI  \\ 
20061217 & 54086.26 & -2485.6 & - &  $>  21.6 $&  - &  - & SI  \\ 
20070207 & 54138.18 & -2433.7 & - &  $>  22.4 $&  - &  - & SI  \\ 
20070313 & 54172.18 & -2399.7 & - &  $>  21.7 $&  - &  - & SI  \\ 
20071016 & 54389.42 & -2182.4 & - &  $>  21.6 $&  - &  - & SI  \\ 
20071215 & 54449.38 & -2122.5 & - &  $>  21.8 $&  - &  - & SI  \\ 
20071231 & 54465.20 & -2106.7 & - &  $>  21.9 $&  - &  - & SI  \\ 
20080112 & 54477.29 & -2094.6 & - &  $>  21.7 $&  - &  - & SI  \\ 
20080207 & 54503.15 & -2068.7 & - &  $>  21.6 $&  - &  - & SI  \\ 
20080312 & 54537.12 & -2034.7 & - &  $>  21.7 $&  - &  - & SI  \\ 
20080404 & 54560.13 & -2011.7 & - &  $>  21.7 $&  - &  - & SI  \\ 
20080928 & 54737.45 & -1834.4 & - &  $>  21.7 $&  - &  - & SI  \\ 
20081130 & 54800.39 & -1771.5 & - &  $>  21.6 $&  - &  - & SI  \\ 
20090120 & 54851.20 & -1720.7 & - &  $>  21.4 $&  - &  - & SI  \\ 
20091017 & 55121.47 & -1450.4 & - &  $>  22.2 $&  - &  - & SI  \\ 
20091109 & 55144.44 & -1427.4 & - &  $>  21.4 $&  - &  - & SI  \\ 
20091111 & 55146.98 & -1424.9 & - &  - &  $>  20.0 $&  - & Apogee  \\ 
20091210 & 55175.20 & -1396.7 & - &  $>  21.6 $&  - &  - & SI  \\ 
20100107 & 55203.22 & -1368.6 & - &  $>  21.7 $&  - &  - & SI  \\ 
20100220 & 55247.12 & -1324.7 & - &  $>  21.6 $&  - &  - & SI  \\ 
20100319 & 55274.15 & -1297.7 & - &  $>  21.8 $&  - &  - & SI  \\ 
20100906 & 55445.37 & -1126.5 & - &  - &  $>  20.6 $&  - & PTF  \\ 
20101020 & 55489.99 & -1081.9 & - &  - &  $>  19.2 $&  - & Apogee  \\ 
20101128 & 55528.31 & -1043.5 & - &  $>  21.5 $&  - &  - & SI  \\ 
20101203 & 55533.33 & -1038.5 & - &  $>  21.9 $&  - &  - & SI  \\ 
20101209 & 55539.98 & -1031.9 & - &  - &  $>  19.5 $&  - & Apogee  \\ 
20101225 & 55555.26 & -1016.6 & - &  $>  21.9 $&  - &  - & SI  \\ 
20110112 & 55573.08 &  -998.8 & - &  $>  21.6 $&  - &  - & SI  \\ 
20110127 & 55588.16 &  -983.7 & - &  $>  21.9 $&  - &  - & SI  \\ 
20110207 & 55599.21 &  -972.6 & - &  - &  $>  20.5 $&  - & PTF  \\ 
20110306 & 55626.12 &  -945.7 & - &  $>  21.6 $&  - &  - & SI  \\ 
20110326 & 55646.14 &  -925.7 & - &  $>  21.1 $&  - &  - & SI  \\ 
20111031 & 55865.11 &  -706.7 & - &  - &  $>  19.4 $&  - & Apogee  \\ 
20111117 & 55882.05 &  -689.8 & - &  - &  $>  19.1 $&  - & Apogee  \\ 
20111225 & 55920.18 &  -651.7 & - &  21.75 (0.35) &  - &  - & SI  \\ 
20120117 & 55943.80 &  -628.0 & - &  - &  $>  19.2 $&  - & Apogee  \\ 
20120129 & 55955.13 &  -616.7 & - &  21.68 (0.30) &  - &  - & SI  \\ 
20120217 & 55974.82 &  -597.0 & - &  - &  $>  19.4 $&  - & Apogee  \\ 
20120226 & 55983.79 &  -588.1 & - &  - &  $>  19.4 $&  - & Apogee  \\ 
20120314 & 56000.16 &  -571.7 & - &  21.79 (0.40) &  - &  - & SI  \\ 
20120323 & 56009.17 &  -562.7 & - &  21.69 (0.47) &  - &  - & SI  \\ 
20120925 & 56195.41 &  -376.4 & - &  20.86 (0.21) &  - &  - & SI  \\ 
20120926 & 56196.42 &  -375.4 & - &  20.81 (0.43) &  - &  - & T21FLI  \\ 
20121023 & 56223.10 &  -348.8 & - &  - &  19.57 (0.46) &  - & Apogee  \\ 
20121107 & 56238.34 &  -333.5 & - &  20.71 (0.19) &  - &  - & SI  \\ 
20121203 & 56264.18 &  -307.7 & - &  20.64 (0.48) &  - &  - & SI  \\ 
20121208 & 56269.31 &  -302.5 & - &  20.59 (0.19) &  - &  - & SI  \\ 
20121211 & 56272.98 &  -298.9 & - &  - &  19.33 (0.22) &  - & Apogee  \\ 
20130106 & 56298.22 &  -273.6 & - &  20.60 (0.20) &  - &  - & SI  \\ 
20130107 & 56303.91 &  -267.9 & - &  - &  $>  18.7 $&  - & Apogee  \\ 
20130113 & 56305.17 &  -266.7 & - &  20.55 (0.21) &  - &  - & SI  \\ 
20130203 & 56326.74 &  -245.1 & - &  - &  19.21 (0.19) &  - & Apogee  \\ 
\hline  
\end{tabular}
\end{table*}

\begin{table*}
\contcaption{}
\begin{tabular}{@{}cccccccc@{}}
\hline  
Date & MJD & Phase$^a$ & $B$ & $V$ & $R$  & $I$ & Instrument Key \\ 
 &  & (days) & (mag) & (mag) & (mag)  & (mag) &  \\ 
\hline  
20130223 & 56346.19 &  -225.7 & - &  20.50 (0.24) &  - &  - & SI  \\ 
20130818 & 56522.14 &   -49.7 & - &  - &  17.93 (0.33) &  - & FLI-New  \\ 
20130819 & 56523.11 &   -48.7 & - &  - &  17.91 (0.28) &  - & FLI-New  \\ 
20130820 & 56524.17 &   -47.7 & - &  - &  17.87 (0.24) &  - & LRS  \\ 
20130821 & 56525.06 &   -46.8 & 19.39 (0.24) &  18.54 (0.13) &  17.72 (0.09) &  17.40 (0.09) & SBIG  \\ 
20130821 & 56525.43 &   -46.4 & 19.35 (0.12) &  18.41 (0.07) &  - &  - & kb74  \\ 
20130822 & 56526.10 &   -45.8 & 19.31 (0.24) &  18.50 (0.20) &  17.67 (0.11) &  17.31 (0.09) & SBIG  \\ 
20130822 & 56526.44 &   -45.4 & 19.34 (0.11) &  18.40 (0.05) &  - &  - & kb74  \\ 
20130824 & 56528.79 &   -43.1 & 19.25 (0.17) &  18.50 (0.08) &  - &  - & kb71  \\ 
20130825 & 56529.40 &   -42.4 & 19.22 (0.09) &  18.41 (0.06) &  - &  - & kb78  \\ 
20130828 & 56532.39 &   -39.5 & 19.02 (0.08) &  18.30 (0.04) &  - &  - & kb73  \\ 
20130901 & 56536.08 &   -35.8 & 18.87 (0.12) &  18.08 (0.06) &  17.29 (0.08) &  16.93 (0.12) & AFOSC  \\ 
20130905 & 56540.16 &   -31.7 & - &  - &  17.39 (0.20) &  - & Apogee  \\ 
20130905 & 56540.39 &   -31.5 & - &  18.04 (0.19) &  - &  - & T21FLI  \\ 
20130906 & 56541.46 &   -30.4 & 18.89 (0.05) &  18.06 (0.04) &  - &  - & kb74  \\ 
20130907 & 56542.44 &   -29.4 & 18.83 (0.06) &  17.99 (0.04) &  - &  - & kb74  \\ 
20130913 & 56548.18 &   -23.7 & 18.68 (0.04) &  17.93 (0.03) &  - &  - & IO:O  \\ 
20130913 & 56548.43 &   -23.4 & 18.69 (0.05) &  17.88 (0.04) &  - &  - & kb74  \\ 
20130914 & 56549.43 &   -22.4 & 18.62 (0.04) &  17.82 (0.03) &  - &  - & kb74  \\ 
20130922 & 56557.12 &   -14.7 & - &  - &  16.94 (0.18) &  - & Apogee  \\ 
20130924 & 56559.08 &   -12.8 & 18.35 (0.03) &  17.56 (0.04) &  - &  - & kb70  \\ 
20130925 & 56560.39 &   -11.5 & 18.32 (0.06) &  17.52 (0.04) &  - &  - & kb74  \\ 
20130926 & 56561.39 &   -10.5 & 18.27 (0.05) &  17.52 (0.04) &  - &  - & kb74  \\ 
20130928 & 56563.33 &    -8.5 & 18.28 (0.04) &  17.49 (0.04) &  - &  - & kb78  \\ 
20131001 & 56566.20 &    -5.7 & 18.22 (0.06) &  17.47 (0.04) &  - &  - & IO:O  \\ 
20131006 & 56571.21 &    -0.6 & 18.23 (0.03) &  17.46 (0.04) &  - &  - & IO:O  \\ 
20131007 & 56572.29 &     0.4 & 18.22 (0.04) &  17.45 (0.04) &  - &  - & kb73  \\ 
20131008 & 56573.46 &     1.6 & - &  17.43 (0.22) &  - &  - & T21FLI  \\ 
20131009 & 56574.06 &     2.2 & 18.22 (0.05) &  17.46 (0.05) &  - &  - & kb70  \\ 
20131010 & 56575.15 &     3.3 & - &  17.41 (0.24) &  - &  - & T18SBIG  \\ 
20131011 & 56576.77 &     4.9 & - &  - &  16.90 (0.18) &  - & T17FLI \\ 
20131012 & 56577.35 &     5.5 & 18.23 (0.05) &  17.43 (0.02) &  - &  - & kb78  \\ 
20131014 & 56579.03 &     7.2 & 18.24 (0.04) &  17.44 (0.04) &  - &  - & kb70  \\ 
20131014 & 56579.10 &     7.2 & - &  17.49 (0.18) &  - &  - & T18SBIG  \\ 
20131017 & 56582.04 &    10.2 & 18.33 (0.06) &  17.53 (0.04) &  - &  - & kb75  \\ 
20131017 & 56582.44 &    10.6 & - &  17.51 (0.18) &  - &  - & T21FLI  \\ 
20131019 & 56584.68 &    12.8 & - &  - &  16.98 (0.21) &  - & T17FLI  \\ 
20131020 & 56585.04 &    13.2 & 18.36 (0.05) &  17.56 (0.04) &  - &  - & kb70  \\ 
20131024 & 56589.68 &    17.8 & - &  - &  17.10 (0.31) &  - & T17FLI  \\ 
20131026 & 56591.35 &    19.5 & 18.61 (0.05) &  17.80 (0.05) &  - &  - & kb73  \\ 
20131026 & 56591.63 &    19.8 & - &  17.86 (0.24) &  - &  - & T9SBIG  \\ 
20131026 & 56591.67 &    19.8 & - &  - &  17.09 (0.16) &  - & T17FLI  \\ 
20131028 & 56593.09 &    21.2 & - &  17.77 (0.13) &  - &  - & T18SBIG  \\ 
20131028 & 56593.38 &    21.5 & - &  17.83 (0.08) &  17.06 (0.05) &  - & T21FLI  \\ 
20131030 & 56595.04 &    23.2 & - &  - &  17.26 (0.16) &  - & T17FLI \\ 
20131030 & 56595.21 &    23.4 & 18.68 (0.04) &  17.89 (0.04) &  - &  - & IO:O  \\ 
20131101 & 56597.08 &    25.2 & - &  - &  17.14 (0.18) &  - & Apogee  \\ 
20131102 & 56598.50 &    26.7 & 18.86 (0.05) &  17.99 (0.04) &  - &  - & kb74  \\ 
20131104 & 56600.99 &    29.1 & 18.89 (0.06) &  18.01 (0.05) &  - &  - & kb75  \\ 
20131104 & 56601.00 &    29.1 & 18.84 (0.05) &  18.04 (0.04) &  - &  - & kb75  \\ 
20131105 & 56601.10 &    29.2 & 18.86 (0.04) &  18.04 (0.04) &  - &  - & IO:O  \\ 
20131107 & 56603.04 &    31.2 & 18.91 (0.04) &  18.05 (0.02) &  - &  - & IO:O  \\ 
20131107 & 56603.10 &    31.2 & 18.84 (0.07) &  18.10 (0.05) &  17.20 (0.07) &  16.80 (0.04) & SBIG  \\ 
20131107 & 56603.42 &    31.6 & - &  18.03 (0.16) &  - &  - & T21FLI \\ 
20131108 & 56604.24 &    32.4 & 18.93 (0.05) &  18.06 (0.03) &  - &  - & kb78  \\ 
20131111 & 56607.25 &    35.4 & 19.01 (0.06) &  18.17 (0.04) &  - &  - & kb77  \\ 
20131111 & 56607.28 &    35.4 & 19.04 (0.06) &  18.18 (0.04) &  - &  - & kb78  \\ 
20131112 & 56608.12 &    36.3 & - &  18.15 (0.21) &  - &  - & T18SBIG  \\ 
20131112 & 56608.89 &    37.0 & 18.92 (0.07) &  18.25 (0.07) &  17.26 (0.07) &  16.89 (0.07) & SBIG  \\ 
20131113 & 56609.01 &    37.2 & - &  18.17 (0.08) &  17.28 (0.07) &  - & T7SBIG  \\ 
\hline  
\end{tabular}
\end{table*}

\begin{table*}
\contcaption{}
\begin{tabular}{@{}cccccccc@{}}
\hline  
Date & MJD & Phase$^a$ & $B$ & $V$ & $R$  & $I$ & Instrument Key \\ 
 &  & (days) & (mag) & (mag) & (mag)  & (mag) &  \\ 
\hline  
20131115 & 56611.08 &    39.2 & 19.06 (0.06) &  18.25 (0.04) &  - &  - & IO:O  \\ 
20131115 & 56611.19 &    39.3 & 19.11 (0.08) &  18.22 (0.04) &  - &  - & kb78  \\ 
20131117 & 56613.25 &    41.4 & 19.25 (0.13) &  18.32 (0.06) &  - &  - & kb78  \\ 
20131120 & 56616.29 &    44.4 & 19.22 (0.11) &  18.49 (0.08) &  - &  - & kb78  \\ 
20131122 & 56618.05 &    46.2 & 19.36 (0.04) &  18.49 (0.03) &  - &  - & IO:O  \\ 
20131122 & 56618.96 &    47.1 & 19.35 (0.07) &  18.57 (0.06) &  - &  - & kb79  \\ 
20131124 & 56620.98 &    49.1 & 19.41 (0.04) &  18.57 (0.03) &  - &  - & IO:O  \\ 
20131126 & 56622.04 &    50.2 & - &  - &  17.66 (0.21) &  - & Apogee  \\ 
20131126 & 56622.99 &    51.1 & 19.48 (0.04) &  18.61 (0.04) &  - &  - & IO:O  \\ 
20131127 & 56623.14 &    51.3 & 19.55 (0.07) &  18.60 (0.06) &  - &  - & kb78  \\ 
20131129 & 56625.04 &    53.2 & 19.50 (0.04) &  18.64 (0.03) &  - &  - & IO:O  \\ 
20131129 & 56625.18 &    53.3 & 19.59 (0.05) &  18.63 (0.04) &  - &  - & kb78  \\ 
20131202 & 56628.54 &    56.7 & 19.57 (0.07) &  18.74 (0.04) &  - &  - & kb05  \\ 
20131202 & 56628.96 &    57.1 & - &  18.70 (0.13) &  17.84 (0.10) &  - & T7SBIG  \\ 
20131203 & 56629.95 &    58.1 & - &  - &  17.99 (0.25) &  - & Apogee  \\ 
20131206 & 56632.00 &    60.2 & - &  - &  - &  17.59 (0.21) & AFOSC  \\ 
20131207 & 56633.00 &    61.2 & - &  18.81 (0.20) &  - &  - & T7SBIG  \\ 
20131208 & 56634.14 &    62.3 & 19.77 (0.07) &  18.86 (0.05) &  - &  - & kb74  \\ 
20131208 & 56634.97 &    63.1 & - &  - &  18.00 (0.24) &  - & Apogee  \\ 
20131210 & 56636.32 &    64.5 & 19.86 (0.06) &  18.89 (0.05) &  - &  - & kb74  \\ 
20131214 & 56640.33 &    68.5 & 19.99 (0.16) &  18.99 (0.13) &  - &  - & kb74  \\ 
20131217 & 56643.97 &    72.1 & - &  - &  18.36 (0.30) &  - & Apogee  \\ 
20131222 & 56648.47 &    76.6 & 20.13 (0.18) &  19.16 (0.07) &  - &  - & kb71  \\ 
20131223 & 56649.76 &    77.9 & - &  19.25 (0.26) &  18.31 (0.12) &  - & T7SBIG  \\ 
20131225 & 56651.92 &    80.1 & 20.21 (0.13) &  19.22 (0.05) &  - &  - & kb75  \\ 
20140101 & 56658.93 &    87.1 & - &  - &  18.66 (0.20) &  - & Apogee  \\ 
20140102 & 56659.97 &    88.1 & 20.38 (0.07) &  19.36 (0.04) &  - &  - & IO:O  \\ 
20140103 & 56660.96 &    89.1 & 20.42 (0.04) &  19.43 (0.04) &  18.50 (0.03) &  18.11 (0.05) & LRS  \\ 
20140104 & 56661.96 &    90.1 & 20.45 (0.04) &  19.40 (0.03) &  - &  - & IO:O  \\ 
20140106 & 56663.91 &    92.1 & - &  - &  18.79 (0.37) &  - & Apogee  \\ 
20140107 & 56664.89 &    93.0 & 20.54 (0.36) &  19.58 (0.09) &  18.59 (0.04) &  18.26 (0.09) & AFOSC  \\ 
20140107 & 56664.93 &    93.1 & 20.55 (0.32) &  19.52 (0.09) &  - &  - & IO:O  \\ 
20140117 & 56674.33 &   102.5 & 20.61 (0.30) &  19.78 (0.28) &  - &  - & kb74  \\ 
20140121 & 56678.84 &   107.0 & 20.73 (0.23) &  19.78 (0.13) &  18.89 (0.47) &  18.42 (0.17) & SBIG  \\ 
20140125 & 56682.74 &   110.9 & 20.75 (0.23) &  19.86 (0.10) &  18.95 (0.05) &  18.68 (0.15) & AFOSC  \\ 
20140126 & 56683.76 &   111.9 & - &  - &  19.11 (0.31) &  - & Apogee  \\ 
20140128 & 56685.45 &   113.6 & 20.99 (0.25) &  19.84 (0.07) &  - &  - & kb71  \\ 
20140128 & 56685.92 &   114.1 & 20.89 (0.09) &  19.98 (0.07) &  - &  - & IO:O  \\ 
20140131 & 56688.44 &   116.6 & 21.18 (0.32) &  20.02 (0.09) &  - &  - & kb71  \\ 
20140202 & 56690.80 &   119.0 & 20.95 (0.42) &  20.20 (0.10) &  - &  - & kb70  \\ 
20140205 & 56693.91 &   122.1 & 21.14 (0.10) &  20.18 (0.05) &  - &  - & IO:O  \\ 
20140206 & 56694.72 &   122.9 & - &  - &  19.14 (0.45) &  - & Apogee  \\ 
20140207 & 56695.43 &   123.6 & 21.05 (0.33) &  20.35 (0.14) &  - &  - & kb05  \\ 
20140208 & 56696.44 &   124.6 & - &  20.20 (0.33) &  - &  - & kb71  \\ 
20140211 & 56699.86 &   128.0 & 21.14 (0.10) &  20.25 (0.09) &  - &  - & IO:O  \\ 
20140214 & 56702.86 &   131.0 & 20.98 (0.25) &  20.25 (0.20) &  - &  - & IO:O  \\ 
20140301 & 56717.86 &   146.0 & 21.79 (0.08) &  20.52 (0.05) &  19.55 (0.05) &  19.07 (0.04) & LRS  \\ 
20140302 & 56718.77 &   146.9 & - &  - &  19.39 (0.28) &  - & Apogee  \\ 
20140330 & 56746.85 &   175.0 & - &  21.02 (0.43) &  20.34 (0.33) &  19.73 (0.33) & AFOSC  \\ 
20140330 & 56746.85 &   175.0 & $>  19.5 $&  - &  - &  - & AFOSC  \\ 
20140805 & 56874.21 &   302.4 & 23.38 (0.17) &  21.96 (0.07) &  21.03 (0.09) &  20.83 (0.09) & ALFOSC  \\ 
20140825 & 56894.20 &   322.3 & 23.24 (0.11) &  22.12 (0.08) &  21.14 (0.06) &  21.01 (0.09) & ALFOSC  \\ 
20140918 & 56918.20 &   346.3 & 23.36 (0.14) &  22.34 (0.10) &  21.37 (0.12) &  21.16 (0.15) & ALFOSC  \\ 
20141013 & 56943.21 &   371.4 & 23.33 (0.27) &  22.55 (0.19) &  21.67 (0.20) &  21.23 (0.25) & ALFOSC  \\ 
20141027 & 56957.27 &   385.4 & 23.54 (0.14) &  22.62 (0.07) &  21.64 (0.10) &  21.27 (0.18) & ALFOSC  \\ 
20141112 & 56973.20 &   401.4 & 23.38 (0.20) &  22.76 (0.13) &  21.64 (0.14) &  21.09 (0.14) & ALFOSC  \\ 
20141124 & 56986.01 &   414.2 & 23.63 (0.12) &  22.63 (0.08) &  21.65 (0.11) &  21.21 (0.13) & ALFOSC  \\ 
20150124 & 57046.87 &   475.0 & - &  - &  22.02 (0.24) &  - & ALFOSC  \\ 
20150310 & 57091.94 &   520.1 & - &  - &  22.11 (0.19) &  - & ALFOSC  \\ 
20150806 & 57240.20 &   668.3 & - &  - &  $>  21.2 $&  - & ALFOSC  \\ 
20161023 & 57684.97 &  1113.1 & - &  23.24 (0.02) &  - &  21.95 (0.01) & WFC3  \\ 
\hline 
\end{tabular}
\begin{flushleft}
$^a$ Phases are relative to $V$ maximum light, MJD $= 56571.8 \pm 1.0$.\\ 
\end{flushleft}
\end{table*}

\begin{table*}
\caption{Optical Sloan photometry of SNhunt151 ({\sc ABmag}).}
\label{table_SLph}
\begin{tabular}{@{}ccccccccc@{}}
\hline  
Date & MJD & Phase$^a$ & $u$ & $g$ & $r$ & $i$  & $z$ & Instrument Key\\ 
 &  & (days) & (mag) & (mag) & (mag) & (mag)  & (mag) &  \\ 
\hline  
20031221 & 52994.50 & -3577.4 & $>  22.5 $&  $>  22.8 $&  $>  22.3 $&  $>  22.1 $&  $>  21.6 $& SDSS  \\ 
20110207 & 55599.14 &  -972.7 & - &  $>  21.5 $&  - &  - &  - & PTF  \\ 
20110208 & 55600.13 &  -971.7 & - &  $>  21.0 $&  - &  - &  - & PTF  \\ 
20130821 & 56525.44 &   -46.4 & - &  18.96 (0.06) &  17.93 (0.05) &  17.66 (0.04) &  - & kb74  \\ 
20130822 & 56526.45 &   -45.4 & - &  18.95 (0.04) &  17.87 (0.04) &  17.64 (0.04) &  - & kb74  \\ 
20130824 & 56528.80 &   -43.0 & - &  18.95 (0.08) &  17.82 (0.05) &  17.64 (0.04) &  - & kb71  \\ 
20130825 & 56529.42 &   -42.4 & - &  18.88 (0.05) &  17.81 (0.04) &  17.60 (0.04) &  - & kb78  \\ 
20130828 & 56532.42 &   -39.4 & - &  - &  17.68 (0.04) &  17.43 (0.02) &  - & kb73  \\ 
20130830 & 56534.17 &   -37.7 & - &  - &  17.62 (0.03) &  17.42 (0.02) &  - & RATCam  \\ 
20130906 & 56541.47 &   -30.4 & - &  18.45 (0.03) &  17.50 (0.04) &  17.29 (0.04) &  - & kb74  \\ 
20130907 & 56542.46 &   -29.4 & - &  18.46 (0.05) &  17.43 (0.04) &  17.27 (0.04) &  - & kb74  \\ 
20130913 & 56548.18 &   -23.7 & 19.29 (0.06) &  - &  17.42 (0.03) &  17.26 (0.02) &  - & IO:O  \\ 
20130914 & 56549.46 &   -22.4 & - &  18.17 (0.03) &  17.28 (0.02) &  17.05 (0.02) &  - & kb74  \\ 
20130924 & 56559.10 &   -12.8 & - &  17.93 (0.04) &  17.00 (0.03) &  16.78 (0.03) &  - & kb70  \\ 
20130925 & 56560.42 &   -11.4 & - &  17.86 (0.03) &  16.95 (0.04) &  16.72 (0.03) &  - & kb74  \\ 
20130926 & 56561.42 &   -10.4 & - &  17.89 (0.02) &  16.98 (0.03) &  16.74 (0.02) &  - & kb74  \\ 
20130928 & 56563.36 &    -8.5 & - &  17.91 (0.02) &  16.93 (0.03) &  16.68 (0.02) &  - & kb78  \\ 
20131001 & 56566.20 &    -5.7 & 18.78 (0.11) &  - &  16.90 (0.03) &  16.73 (0.03) &  - & IO:O  \\ 
20131006 & 56571.21 &    -0.6 & 18.83 (0.05) &  17.84 (0.02) &  16.87 (0.03) &  16.72 (0.02) &  - & IO:O  \\ 
20131007 & 56572.30 &     0.5 & - &  17.81 (0.05) &  16.86 (0.04) &  16.67 (0.04) &  - & kb73  \\ 
20131009 & 56574.07 &     2.2 & - &  17.79 (0.03) &  16.87 (0.03) &  16.68 (0.03) &  - & kb70  \\ 
20131012 & 56577.36 &     5.5 & - &  17.79 (0.02) &  16.86 (0.03) &  16.64 (0.02) &  - & kb78  \\ 
20131014 & 56579.04 &     7.2 & - &  17.84 (0.03) &  16.91 (0.03) &  16.66 (0.03) &  - & kb70  \\ 
20131017 & 56582.05 &    10.2 & - &  17.90 (0.03) &  16.95 (0.02) &  16.70 (0.03) &  - & kb75  \\ 
20131020 & 56585.07 &    13.2 & - &  17.91 (0.12) &  - &  16.76 (0.16) &  - & kb70  \\ 
20131026 & 56591.32 &    19.5 & - &  18.18 (0.04) &  17.19 (0.02) &  16.97 (0.02) &  16.79 (0.06) & kb73  \\ 
20131030 & 56595.21 &    23.4 & 19.49 (0.12) &  18.31 (0.03) &  17.22 (0.03) &  17.09 (0.03) &  - & IO:O  \\ 
20131102 & 56598.68 &    26.8 & - &  18.44 (0.03) &  17.29 (0.03) &  17.17 (0.03) &  16.90 (0.05) & kb71  \\ 
20131105 & 56601.10 &    29.2 & 19.70 (0.10) &  18.47 (0.03) &  17.32 (0.02) &  17.24 (0.03) &  - & IO:O  \\ 
20131105 & 56601.24 &    29.4 & - &  18.48 (0.03) &  17.34 (0.05) &  17.21 (0.03) &  16.99 (0.05) & kb73  \\ 
20131107 & 56603.04 &    31.2 & 19.73 (0.12) &  18.52 (0.03) &  17.34 (0.03) &  17.27 (0.04) &  - & IO:O  \\ 
20131108 & 56604.30 &    32.5 & - &  18.56 (0.03) &  17.38 (0.03) &  17.24 (0.02) &  17.02 (0.04) & kb78  \\ 
20131111 & 56607.26 &    35.4 & - &  18.60 (0.02) &  17.48 (0.02) &  17.30 (0.04) &  17.09 (0.07) & kb77  \\ 
20131112 & 56608.43 &    36.6 & - &  18.65 (0.03) &  17.48 (0.03) &  17.33 (0.03) &  17.12 (0.06) & kb74  \\ 
20131115 & 56611.08 &    39.2 & 20.12 (0.15) &  18.64 (0.04) &  17.53 (0.03) &  17.46 (0.03) &  - & IO:O  \\ 
20131116 & 56612.25 &    40.4 & - &  18.75 (0.03) &  17.55 (0.03) &  17.45 (0.02) &  - & kb77  \\ 
20131117 & 56613.21 &    41.4 & - &  18.67 (0.12) &  - &  17.52 (0.22) &  17.29 (0.10) & kb77  \\ 
20131120 & 56616.22 &    44.4 & - &  18.82 (0.08) &  17.67 (0.04) &  17.57 (0.05) &  17.34 (0.08) & kb73  \\ 
20131122 & 56618.04 &    46.2 & 20.39 (0.09) &  18.93 (0.02) &  17.73 (0.02) &  17.70 (0.02) &  - & IO:O  \\ 
20131123 & 56619.16 &    47.3 & - &  18.96 (0.04) &  17.69 (0.03) &  17.76 (0.04) &  17.46 (0.05) & kb73  \\ 
20131124 & 56620.97 &    49.1 & 20.45 (0.06) &  18.95 (0.03) &  17.76 (0.03) &  17.75 (0.03) &  - & IO:O  \\ 
20131126 & 56622.23 &    50.4 & - &  19.02 (0.04) &  17.80 (0.04) &  17.77 (0.03) &  17.59 (0.06) & kb77  \\ 
20131126 & 56622.99 &    51.1 & 20.52 (0.10) &  19.07 (0.03) &  17.81 (0.02) &  17.76 (0.03) &  - & IO:O  \\ 
20131129 & 56625.04 &    53.2 & 20.56 (0.06) &  19.11 (0.03) &  17.86 (0.03) &  17.83 (0.03) &  - & IO:O  \\ 
20131130 & 56626.20 &    54.3 & - &  19.13 (0.03) &  17.86 (0.03) &  17.80 (0.02) &  17.67 (0.06) & kb78  \\ 
20131202 & 56628.56 &    56.7 & - &  19.23 (0.04) &  17.95 (0.03) &  17.86 (0.04) &  17.73 (0.10) & kb05  \\ 
20131214 & 56640.15 &    68.3 & - &  19.54 (0.09) &  18.30 (0.05) &  18.22 (0.04) &  18.04 (0.09) & kb77  \\ 
20131222 & 56648.08 &    76.2 & - &  19.71 (0.04) &  18.41 (0.04) &  18.36 (0.04) &  - & kb77  \\ 
20131225 & 56651.88 &    80.0 & - &  19.76 (0.03) &  18.46 (0.04) &  18.41 (0.04) &  - & kb75  \\ 
20140102 & 56659.96 &    88.1 & 21.47 (0.18) &  19.94 (0.03) &  18.54 (0.03) &  18.56 (0.04) &  - & IO:O  \\ 
20140104 & 56661.96 &    90.1 & 21.54 (0.15) &  19.99 (0.03) &  18.58 (0.02) &  18.59 (0.02) &  - & IO:O  \\ 
20140107 & 56664.93 &    93.1 & - &  20.04 (0.06) &  18.63 (0.03) &  18.67 (0.04) &  - & IO:O  \\ 
20140116 & 56673.34 &   101.5 & - &  20.29 (0.33) &  18.70 (0.10) &  18.77 (0.17) &  - & kb74  \\ 
20140121 & 56678.85 &   107.0 & - &  - &  18.97 (0.04) &  18.89 (0.04) &  - & kb70  \\ 
20140128 & 56685.43 &   113.6 & - &  - &  19.06 (0.04) &  19.05 (0.05) &  - & kb71  \\ 
20140128 & 56685.92 &   114.1 & 21.89 (0.34) &  20.54 (0.07) &  19.11 (0.06) &  19.08 (0.16) &  - & IO:O  \\ 
20140204 & 56692.81 &   121.0 & - &  - &  19.20 (0.09) &  19.27 (0.20) &  - & kb76  \\ 
20140205 & 56693.91 &   122.1 & 22.25 (0.35) &  20.78 (0.06) &  19.23 (0.03) &  19.28 (0.03) &  - & IO:O  \\ 
20140207 & 56695.45 &   123.6 & - &  - &  19.24 (0.07) &  19.31 (0.07) &  - & kb71  \\ 
20140208 & 56696.42 &   124.6 & - &  - &  19.28 (0.10) &  19.38 (0.19) &  - & kb05  \\ 
\hline  
\end{tabular}
\end{table*}

\begin{table*}
\contcaption{}
\begin{tabular}{@{}ccccccccc@{}}
\hline 
Date & MJD & Phase$^a$ & $u$ & $g$ & $r$ & $i$  & $z$ & Instrument Key\\ 
 &  & (days) & (mag) & (mag) & (mag) & (mag)  & (mag) &  \\ 
\hline 
20140211 & 56699.86 &   128.0 & - &  20.83 (0.07) &  19.36 (0.04) &  19.42 (0.07) &  - & IO:O  \\ 
20140214 & 56702.86 &   131.0 & - &  20.75 (0.19) &  19.38 (0.07) &  19.44 (0.08) &  - & IO:O  \\ 
20140215 & 56703.21 &   131.4 & - &  - &  19.42 (0.22) &  19.27 (0.32) &  - & kb74  \\ 
20140301 & 56717.86 &   146.0 & - &  21.21 (0.04) &  19.71 (0.03) &  19.69 (0.06) &  - & LRS  \\ 
20140821 & 56890.19 &   318.3 & - &  - &  21.59 (0.17) &  - &  - & OSIRIS  \\ 
20140824 & 56893.21 &   321.4 & - &  22.66 (0.06) &  21.37 (0.11) &  21.12 (0.11) &  - & OSIRIS  \\ 
20141115 & 56976.15 &   404.3 & - &  23.03 (0.25) &  - &  - &  - & OSIRIS  \\ 
20141124 & 56985.21 &   413.4 & - &  $>  22.2 $ &  22.39 (0.22) &  22.33 (0.44) &  - & OSIRIS  \\ 
20161207 & 57729.02 &  1157.2 & - &  - &  $>  22.2 $&  - &  - & AFOSC  \\ 
20170123 & 57776.91 &  1205.1 & - &  - &  $>  21.6 $&  - &  - & AFOSC  \\ 
20170216 & 57800.96 &  1229.1 & - &  24.36 (0.19) &  23.26 (0.10) &  23.17 (0.12) &  - & OSIRIS  \\ 
\hline  
\end{tabular}
\begin{flushleft}
$^a$ Phases are relative to $V$ maximum light, MJD = 56571.8 $\pm$ 1.0.\\ 
\end{flushleft}
\end{table*}

\begin{table*}
\caption{Near-infrared photometry of SNhunt151 ({\sc Vegamag}).}
\label{table_NIRph}
\begin{tabular}{@{}ccccccc@{}}
\hline  
Date & MJD & Phase$^a$ & $J$ & $H$ & $K$ & Instrument Key\\ 
 &  & (days) & (mag) & (mag)  & (mag) &  \\ 
\hline  
20131105 & 56600.08 &    28.2 & 15.77 (0.28) &  15.48 (0.20) &  15.16 (0.28) & NICS  \\ 
20131119 & 56614.06 &    42.2 & 16.09 (0.26) &  15.81 (0.26) &  15.24 (0.28) & NICS  \\ 
20131125 & 56621.04 &    49.2 & 16.18 (0.27) &  16.00 (0.19) &  15.39 (0.31) & NOTCam  \\ 
20131127 & 56622.15 &    50.3 & 16.24 (0.14) &  16.02 (0.22) &  15.45 (0.31) & NICS  \\ 
20131217 & 56643.07 &    71.2 & 16.85 (0.38) &  16.44 (0.22) &  15.94 (0.56) & NOTCam  \\ 
20131223 & 56648.97 &    77.1 & 16.96 (0.32) &  16.59 (0.26) &  16.09 (0.27) & NICS  \\ 
20140116 & 56673.94 &   102.1 & - &  - &  16.66 (0.23) & NOTCam  \\ 
20140212 & 56700.93 &   129.1 & 17.85 (0.29) &  17.47 (0.27) &  17.07 (0.28) & NOTCam  \\ 
20140314 & 56730.87 &   159.0 & 18.24 (0.17) &  17.98 (0.20) &  17.35 (0.23) & NOTCam  \\ 
\hline  
\end{tabular}
\begin{flushleft}
$^a$ Phases are relative to $V$ maximum light, MJD $= 56571.8 \pm 1.0$.\\ 
\end{flushleft}
\end{table*}

\begin{table*}
 \centering
  \caption{Log of Spectroscopic Observations of SNhunt151.}
  \label{table_spec}
  \begin{tabular}{@{}ccccccc@{}}
  \hline
  Date & MJD & Phase$^a$ & Instrumental Setup & Grism/Grating+Slit & Spectral Range & Resolution \\
 &  & & & & (\AA) & (\AA) \\
 \hline
20130820 & 56524.20 & $-47.6$ & TNG+LRS & LR-B+$1.00''$ & 3750--7850 & 12\\
20130902 & 56537.18 & $-34.7$ & TNG+LRS & LR-B+$1.00''$ & 3750--7850 & 12\\
20130914 & 56549.01 & $-22.8$ &  Ekar1.82m+AFOSC & gm4+VPH6+$1.69''$ & 3800--8500 & 13.5/15\\  
20131010 & 56575.07 & 3.2 &  TNG+LRS & LR-B+$1.00''$ & 3750--7800 & 12 \\
20131019 & 56584.31 & 12.5 &  SOAR+Goodman & KOSI 600+$0.84''$ & 6300--8650 & 5\\
20131102 & 56598.05 & 26.2 &  GTC+OSIRIS & R1000B+$1.00''$ & 3600--7750 & 7 \\
20131107 & 56603.26 & 31.4 &  GTC+OSIRIS & R1000B+$1.00''$ & 3600--7750 & 7\\
20131109 & 56605.39 & 33.5 &  Palomar+DBSP & BLUE (600/4000)+$1.00''$ & 4000--10000 & 3\\
20131112 & 56608.04 & 36.2 &  NOT+ALFOSC & gm4+$1.00''$ & 3700--8880 & 14 \\
20131205 & 56631.06 & 59.2 &  Ekar1.82m+AFOSC & gm4+VPH6+$1.69''$ & 3900--8800 & 14\\
20131211 & 56637.00 & 65.2 &  Ekar1.82m+AFOSC  & gm4+VPH6+$1.69''$ & 3900--8800 & 13.5/15\\
20131227 & 56653.75 & 82.0 &  Ekar1.82m+AFOSC & gm4+VPH6+$1.69''$ & 3900--8800 & 13.5/15\\
20140103 & 56660.77 & 89.0 &  TNG+LRS & LR-B/R+$1.00''$ & 3800--9150 & 10.5\\
20140301 & 56717.78 & 146.0 &  TNG+LRS & LR-B+$1.00''$ & 3900--8000 & 10.5 \\
20140821 & 56890.19 & 318.3 &  GTC+OSIRIS & R1000B+$1.00''$ & 3850--7750 & 7\\
20141114 & 56976.16 & 404.3 &  GTC+OSIRIS & R1000B+$1.00''$ & 3850--7750 & 7 \\   
\hline
20131105 & 56601.03 & 29.2 & TNG+NICS & IJ+$1.50''$ & 8750--13,500 & 28\\
20131118 & 56615.00 & 43.1 & TNG+NICS & IJ+$1.50''$ & 8750--13,500 & 28\\
20131224 & 56650.84 & 79.0 & TNG+NICS & IJ+$1.50''$ & 8700--14,500 & 28\\
20131225 & 56651.89 & 80.0 & TNG+NICS & HK+$1.50''$ & 14,750--24,800 & 54\\
\hline
\end{tabular}
\begin{flushleft}
$^a$ Phases are relative to $V$ maximum light, MJD $= 56571.8 \pm 1.0$.\\ 
\end{flushleft}
\end{table*}


\bsp	
\label{lastpage}
\end{document}